\documentclass[aps,prb,amsmath,amssymb,superscriptaddress,preprint,floatfix]{revtex4-2}
\usepackage{graphicx}
\usepackage{dcolumn}
\usepackage{epsfig}
\usepackage{float}

\usepackage[bbgreekl]{mathbbol}
\usepackage{mathrsfs}
\usepackage{slashed}
\usepackage[cal=boondox,scr=boondoxo]{mathalfa}
\usepackage{trsym}
\usepackage{cancel}
\usepackage{subfigure}
\usepackage{units}
\usepackage{upgreek}
\usepackage{color}
\usepackage{mathtools}
\usepackage{graphicx}
\usepackage[mathscr]{euscript}

\newcommand{\lsim}{\mathrel{\rlap{\lower4pt\hbox{\hskip1pt$\sim$}}
   \raise1pt\hbox{$<$}}}
\newcommand{\sss}{s}
\newcommand{\etal}{{\it et al.}}

\newcommand{\beq}{\begin{equation}}
\newcommand{\eeq}{\end{equation}}

\newcommand\vex[1]{\mathbf{#1}}

\usepackage[T3,T1]{fontenc}
\DeclareSymbolFont{tipa}{T3}{cmr}{m}{n}
\DeclareMathAccent{\invbreve}{\mathalpha}{tipa}{16}

\makeatother

\graphicspath{{./figures/pdf/}}

\begin{document}
	
\title{Physical Interpretation of Large Lorentz Violation\\
via Weyl Semimetals}
	
\author{V.\ Alan Kosteleck\'y}
\affiliation{Department of Physics, Indiana University, 
Bloomington, Indiana 47405, USA}
\affiliation{Indiana University Center for Spacetime Symmetries, 
Bloomington, Indiana 47405, USA}
	
\author{Ralf Lehnert}
\affiliation{Department of Physics, Indiana University, 
Bloomington, Indiana 47405, USA}
\affiliation{Indiana University Center for Spacetime Symmetries, 
Bloomington, Indiana 47405, USA}
	
\author{Marco Schreck}
\affiliation{Departamento de F\'isica, Universidade Federal do Maranh\~ao
Campus Universit\'ario do Bacanga, S\~ao Lu\'is (MA), 65085-580, Brazil}
	
\author{Babak Seradjeh}
\affiliation{Department of Physics, Indiana University, 
Bloomington, Indiana 47405, USA}
\affiliation{Indiana University Center for Spacetime Symmetries, 
Bloomington, Indiana 47405, USA}
\affiliation{Quantum Science and Engineering Center, 
Indiana University, Bloomington, Indiana 47405, USA}
	
\date{November 2024}

\begin{abstract}
The physical intepretation of effective field theories 
of fundamental interactions incorporating large Lorentz violation
is a long-standing challenge, 
known as the concordance problem.
In condensed-matter physics,
certain Weyl semimetals with emergent Lorentz invariance
exhibit large Lorentz violation,
thereby offering prospective laboratory analogues
for exploration of this issue.
We take advantage of the mathematical equivalence
between the descriptions of large Lorentz violation
in fundamental and condensed-matter physics
to investigate the primary aspects of the concordance problem,
which arise when the coefficients for Lorentz violation are large
or the observer frame is highly boosted.
Using thermodynamic arguments,
we present a physical solution to the concordance problem
and explore some implications.
\end{abstract}
{
\let\clearpage\relax
\maketitle
}

\newpage

\section{Introduction}
\label{Introduction}

Lorentz invariance,
which involves symmetry under rotations and boosts,
is the basis for relativity 
and is a central assumption in our current best description
of nature at the fundamental level.
Recent years have seen an explosion of interest 
in theoretical and experimental studies of Lorentz violation,
driven by the possibility that tiny observable deviations
from the laws of relativity could provide measurable signals 
from an underlying unified theory of fundamental physics at the Planck scale,
such as strings~\cite{ks89,kp91}. 
Effective field theory~\cite{sw}
provides a powerful theoretical tool 
describing the corresponding effects while incorporating known physics.
In effective field theory,
Lorentz-violating terms are controlled by coefficients
that govern the sizes and features of observable effects.
Provided these coefficients for Lorentz violation are sufficiently small,
the deviations from known physics can be handled perturbatively
in a consistent way.
These ideas form the basis for the extensive recent experimental exploration
of many different types of prospective Lorentz-violating effects 
at sensitivities reaching Planck-scale suppression and beyond~\cite{tables}
and are reviewed, for example, in Refs.~\cite{r1,r2,r3,r4,r5,r6}.

The success of the approach to Lorentz violation 
based on effective field theory 
raises imperative conceptual issues about its physical interpretation
in regimes where the Lorentz violation cannot be treated perturbatively.
These regimes can arise 
when the coefficients for Lorentz violation are large in a specified frame 
or when the coefficients are small in one frame
but the system is observed from a highly boosted frame.
In these regimes,
features such as negative energies can appear,
so the stability of the physics becomes subject to question.
Establishing a consistent framework for handling large Lorentz violation
that also incorporates highly boosted observers 
is known as the concordance problem.
Treatments of physical Lorentz violation in fundamental theories to date 
typically sidestep the concordance problem
by assuming small coefficients for Lorentz violation
and restricting allowed boosts to a set of concordant frames
in which the effects remain perturbative~\cite{kl01}.
However,
the lack of a satisfactory resolution to the concordance problem
has left open to question the intrinsic self consistency 
of Lorentz violation in fundamental physics.

In this work,
we take advantage of a close parallel between
the description of Weyl semimetals 
and the effective field theory of Lorentz-violating fermions 
to tackle these issues.
A Weyl semimetal has a band structure
with nodal features such as isolated touching points
near which the dynamics of a quasiparticle excitation
exhibits experimentally observable deviations 
from an emergent (3+1)-dimensional 
Lorentz invariance~\cite{lv15,at16,yf17,amv18,gvkr19}.
Recently, 
the general approach to Lorentz violation in fundamental physics
using effective field theory~\cite{kp95,ck97,ck98,ak04,kl21}
has been applied to characterize types of Weyl semimetals
and their properties~\cite{klmss22,gdmu24}.
In what follows we reverse this direction of reasoning,
demonstrating that the physical existence of Weyl semimetals
and the associated quasiparticle dynamics near the Weyl points 
can be exploited to address the concordance problem.
A key observation is that in the physical semimetal context
no restriction on the size of Lorentz violation exists {\it a priori},
even in the limit of small excitation energies.
This provides a strong indication that a corresponding solution
to the concordance problem must exist 
for certain fundamental particle theories with Lorentz violation,
including in the low-energy limit.
Here,
we use this observation to develop a physical interpretation 
of a model with large Lorentz violation,
address the concordance problem,
and outline some physical implications.

The organization of this paper is as follows.
Section \ref{The concordance problem} provides background material
for our treatment.
A model with large Lorentz violation
applicable in both the semimetal and the fundamental-physics contexts
is presented in Sec.~\ref{Model}.
The concordance problem and its manifestation in the model
are described in 
Sec.~\ref{The concordance problem subsection}.
Our resolution of the concordance problem
is discussed in Sec.~\ref{Physical resolution}.
Adopting a thermodynamic approach,
we examine the case of Weyl semimetals in Sec.~\ref{Weyl semimetals}
and then extend the ideas to the fundamental-particle scenario 
in Sec.~\ref{Fundamental physics}.
Implications of the resolution are discussed in Sec.~\ref{Implications}.
Section \ref{General properties of the vacuum}
presents general features,
while Sec.~\ref{Topological properties of the vacuum}
focuses on topological aspects
and Sec.~\ref{Physical implications}
considers prospective physical observables.
We summarize in Sec.~\ref{Summary}.
Two appendices provide some additional details.
Appendix \ref{Existing approaches}
outlines some aspects of the concordance problem,
while Appendix \ref{Explicit model}
studies a specific model for the thermodynamic bath.
Throughout this work, 
we adopt the conventions of Ref.~\cite{ck97}.

\section{The concordance problem}
\label{The concordance problem}

In the context of fundamental physics,
a model-independent framework based on effective field theory
offers a powerful and widely adopted approach
to describe realistic Lorentz violation~\cite{kp95,ck97,ck98,ak04,kl21}.
In the corresponding Lagrange density,
a given term with Lorentz violation is constructed
by contracting a Lorentz-violating operator with a controlling coefficient. 
The construction ensures invariance of the physics
under general coordinate transformations,
or equivalently under any change of observer frame.
A term in the Lagrange density can be classified 
via the mass dimension of the associated Lorentz-violating operator,
with lower-dimensional operators expected to manifest larger effects
at low energies.

The present work describes a physical resolution of the concordance problem
in the context of a popular model with large Lorentz violation
containing one particular type of Lorentz-violating operator
of low mass dimension.
While the manifestations of the concordance problem
depend partially on the operator~\cite{kl01},
the ideas and results obtained here
are expected to translate to other models,
which would be interesting subjects for future work.
In this section,
we first present the model and its relevance for Weyl semimetals,
and then we outline the associated concordance problem.

\subsection{Model}
\label{Model}

The model we consider in this work is a widely studied theory 
describing the behavior in Minkowski spacetime
of a Dirac fermion $\psi$ of mass $m$
in the presence of a coefficient $b_\mu$ for Lorentz violation.
The Lagrange density for the free theory is~\cite{ck97}
\begin{equation}
\mathcal{L}_b =
\tfrac{1}{2}\overline{\psi}
(\mathrm{i}\cancel{\partial} - m - b_{\mu}\gamma_5\gamma^{\mu})
\psi + \text{h.c.} \, 
\label{lag}
\end{equation}
in natural units.
The components of $b_\mu$ transform as 
a covariant 4-vector under changes of observer frame
but behave as scalars under fermion boosts in a fixed frame,
thereby introducing physical Lorentz violation
and breaking the spin degeneracy of the conventional Dirac theory.
The Dirac equation for this theory can be studied
in the context of relativistic quantum mechanics,
and the Lagrange density (\ref{lag}) can be treated 
as a quantum field theory~\cite{ck97}.
An analysis using Dirac quantization~\cite{pd64}
demonstrates the formal construction and characterization
of the corresponding Fock space~\cite{klss24-1}.

In the theory~(\ref{lag}),
the Lorentz violation is perturbative in concordant frames 
where the components of the coefficient $b_\mu$ are small, 
$|b_\mu|\lsim m$.
We can thus assure conditions suitable for investigating
the concordance problem
by studying the ultrarelativistic limit $m\to 0$,
for which any nonzero $b_\mu$ 
represents large Lorentz violation.
This limit therefore represents a scenario of particular interest
for the analysis to follow.

The theory~(\ref{lag}) has been extensively adopted 
in atomic, nuclear, and particle physics and in astrophysics
as a phenomenological model for Lorentz violation
in electrons, protons, neutrons, 
and other fermions~\cite{bclms24,p1,p2,p3,p4,p5,p6,p7,p8,p9,%
bkl16,ms17,p10,p16,p11,p12,p13,p14,p15,p17,p18,p19,%
p20,p21,p22,p23,p24,p25,p26,p27,p28,p29,%
p30,p31,p32,p33,p34,p35,p36,p37}.
The more formal aspects of the theory have also been investigated 
in the contexts of fundamental quantum field theory%
~\cite{ck97,ck98,jk99,mp99,co99,jc99,jc992,klp02,ba04,ba042,bk05,rl06,%
bt1,bt2,bt3,km13,bt4,bt5,rs17,bt6,bt7,bt9,pv23,bt10}.
The theory also has intriguing geometric properties 
governed by Finsler geometry~\cite{ak11,f1,js13,f2,f3,cm15,f4,f5,f6}.

On the condensed-matter side,
the theory~(\ref{lag}) is known to describe the band structures
of certain semimetals near Weyl nodes%
~\cite{klmss22,gdmu24,ag12,zwb12,s1,kk16,rj16,gc16,kw17,jw17,nx17,%
bskrg19,lqd21,s2,s3,s4}.
In particular,
interactions violating parity inversion and time reversal 
can produce a band structure containing two Weyl cones with nodes 
separated in energy and 3-momentum by $2b_\mu$.
In the vicinity of these Weyl nodes,
the band structure is determined by the dispersion relation 
of the theory~(\ref{lag}) in the limit $m\to 0$~\cite{ck97},
\begin{equation}
(\lambda^2-b^2)^2+4b^2\lambda^2-4(b\cdot \lambda)^2=0\, ,
\label{disp}
\end{equation}
where $\lambda_\mu$ is the wave 4-vector.
In the ground state,
this band structure is filled below the Fermi surface,
and an excitation or hole 
obeys the dispersion relation derived from the theory~(\ref{lag})
with the Fermi velocity in the semimetal
playing the role of the speed of light.

\begin{figure}[t!]
\subfigure[]{\label{fig1a}\includegraphics[width=1.6in]{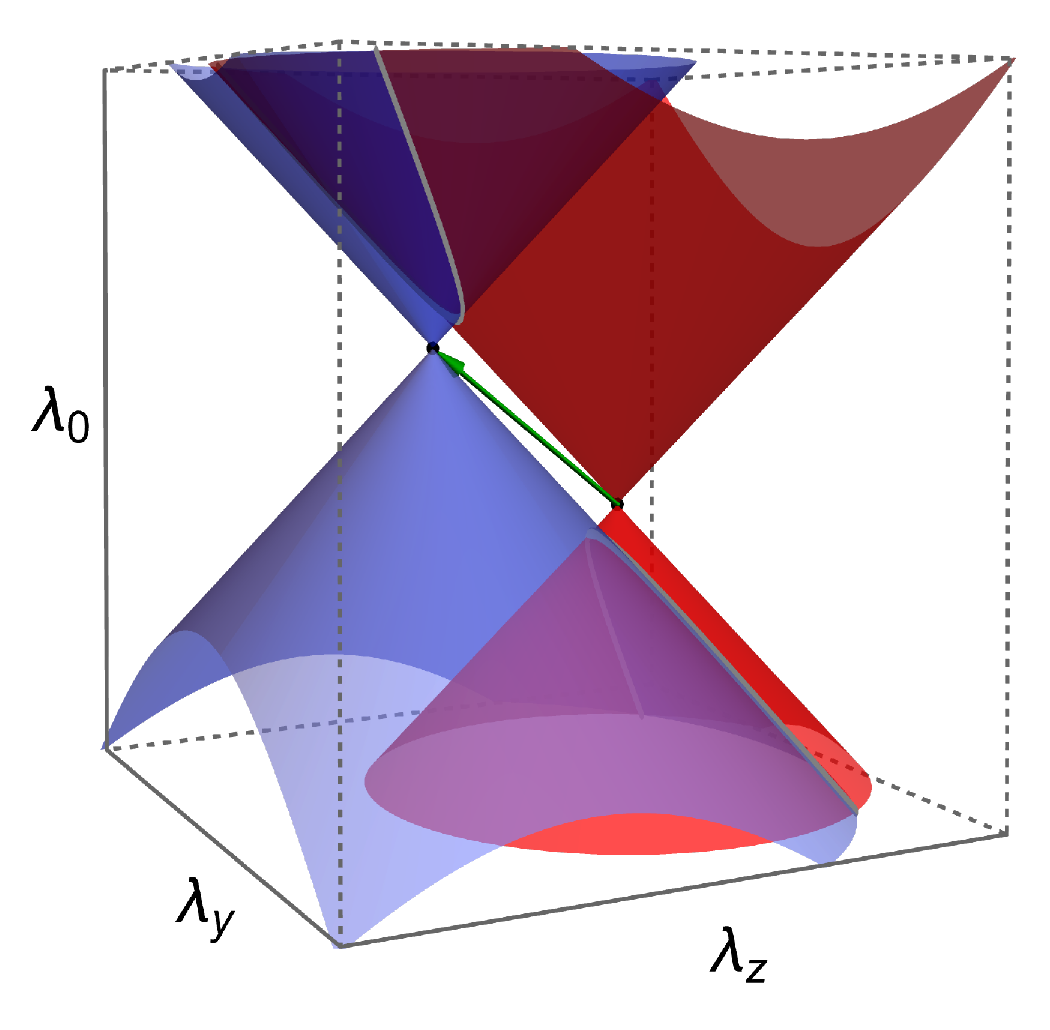}}
\subfigure[]{\label{fig1b}\includegraphics[width=1.6in]{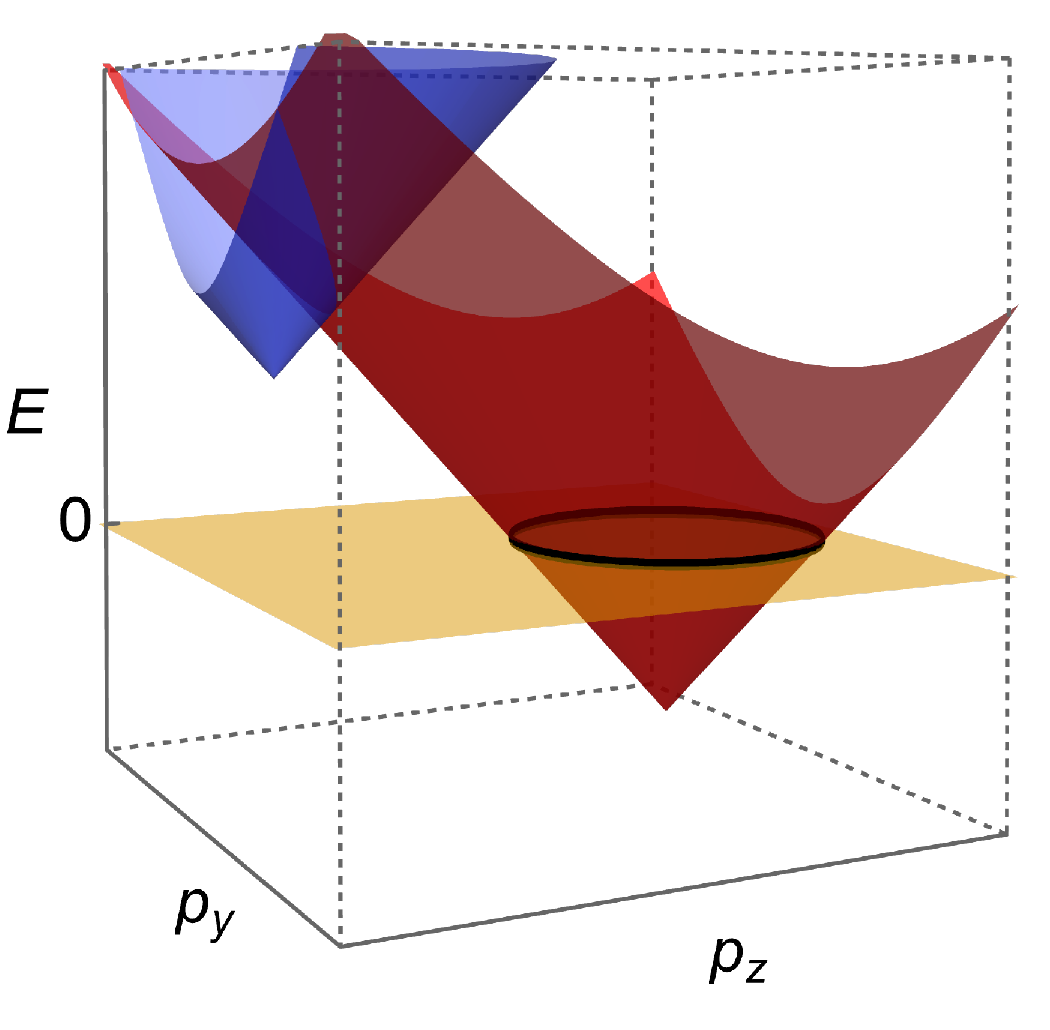}}
\caption{Band structures.}
\label{fig1}
\end{figure}

Figure \ref{fig1a} presents this band structure for zero $x$ component
of the wave vector, 
$\lambda_x=0$,
in the vicinity of the Weyl nodes
in the single-particle description
for the case of spacelike $2b_\mu = (3,0,0,-4)$ in the corresponding units,
indicated by an arrow.
In the many-body field-theoretic picture,
the momentum of the fermion is $p_\mu = \lambda_\mu$,
while that of the antifermion is $p_\mu = -\lambda_\mu$,
and the band structure is mapped into energy levels 
for particle and antiparticle excitations
as displayed in Fig.~\ref{fig1b}.
One part of the band structure,
shown in Fig.~\ref{fig1b} as a cone 
below the planar Fermi surface at zero energy,
represents particle and antiparticle states of negative energy.

\subsection{The concordance problem}
\label{The concordance problem subsection}

For the model (\ref{lag}) with finite $m$,
the concordance problem can manifest itself in two aspects.
First,
in a given observer frame,
the coefficient $b_\mu$ for Lorentz violation
may have components large compared to the mass $m$,
which can produce negative energies 
for particles and antiparticles in the quantum field theory.
The second aspect arises because $b_\mu$ transforms under observer boosts.
As a result,
even if the components of $b_\mu$ 
are small enough in the original observer frame
that all particle and antiparticle energies remain positive,
a sufficiently large observer boost 
generates $b_\mu$ components large compared to the boost-invariant $m$.
This causes negative-energy states for particles and antiparticles
to appear in the boosted frame.
In traditional treatments of quantum field theory,
particle and antiparticle states of negative energy are suspect
and may imply instability of the usual vacuum
and hence instability of the theory.
Some of these issues 
are outlined in Appendix \ref{Existing approaches}.
Resolving the concordance problem in this model 
thus amounts to identifying a consistent treatment and interpretation 
of the appearance of negative energies in both scenarios.

\begin{figure}
\subfigure[]{\label{fig2a}\includegraphics[width=1.6in]{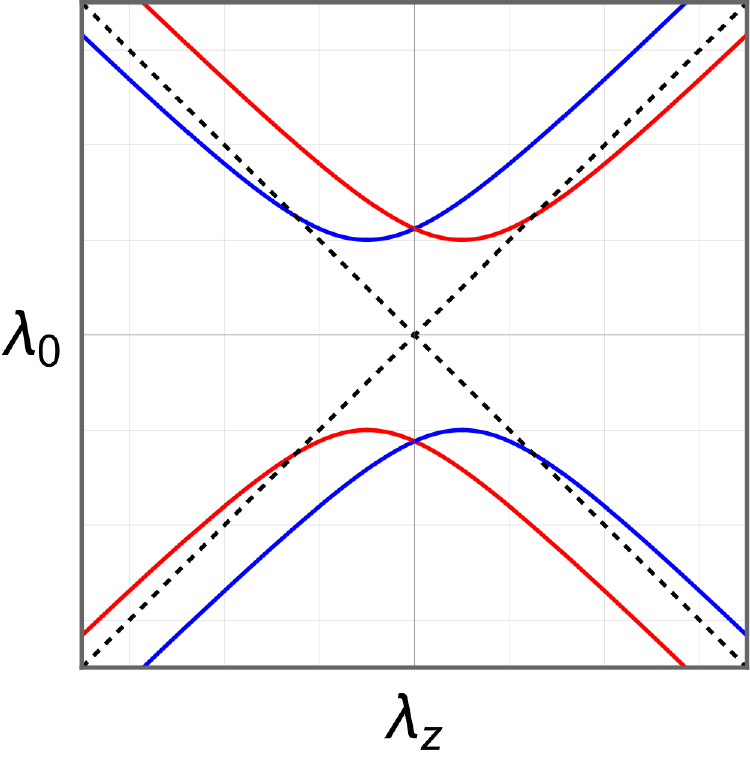}}
\subfigure[]{\label{fig2b}\includegraphics[width=1.6in]{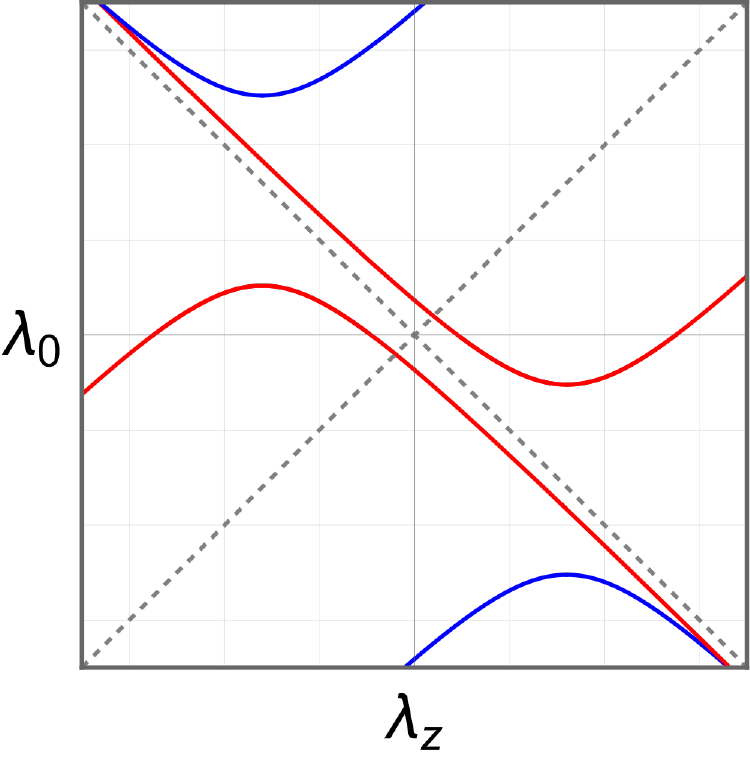}}
\caption{Dispersions for timelike $b_{\mu}$.}
\label{fig2}
\end{figure}

To illustrate the boost issues more explicitly,
consider the model (\ref{lag}) with finite $m$
and choose a definite observer frame.
Suppose first that in this frame $b_\mu$ is purely timelike and perturbative,
in the sense that $|b_0| < m$.
For this case and with $\lambda_x = \lambda_y = 0$,
the four dispersion branches in relativistic quantum mechanics
are plotted in Fig.~\ref{fig2a}.
The perturbative assumption ensures 
a clean separation between the positive- and negative-energy branches
in this frame,
so the ensuing reinterpretation in quantum field theory
generates only positive-energy particles and antiparticles.
The concordance problem therefore has no direct manifestation in this frame.
However,
under an observer boost along the $z$ direction,
the magnitude of the timelike component of the coefficient $b_\mu$ grows,
exceeding $m$ for a sufficiently large boost.
Two of the dispersion branches then cross the $\lambda_z$ axis
in the new frame,
as shown in Fig.~\ref{fig2b}.
The field-theory reinterpretation therefore generates
particle and antiparticle states of negative energy,
a signature of the concordance problem.

\begin{figure}
\subfigure[]{\label{fig3a}\includegraphics[width=1.6in]{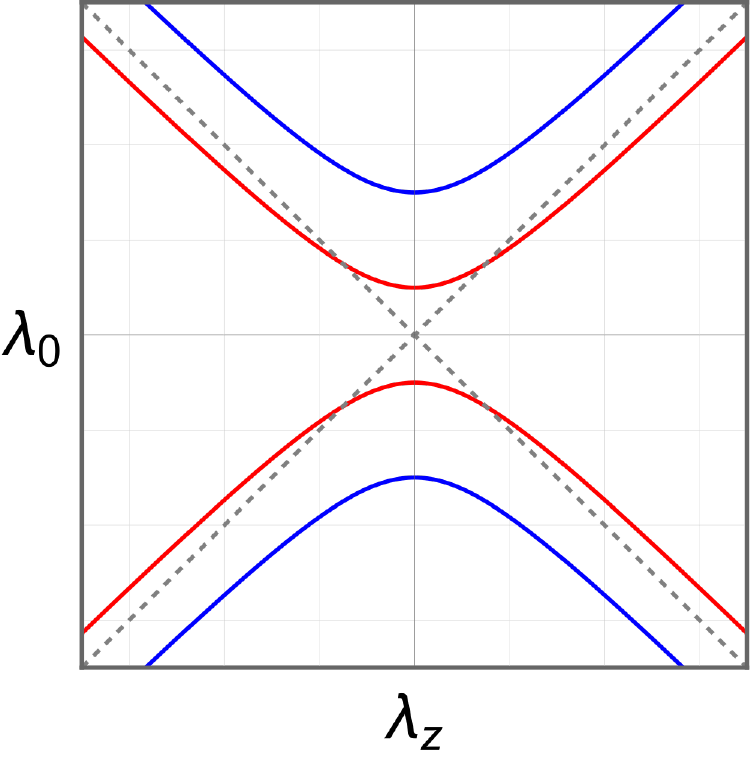}}
\subfigure[]{\label{fig3b}\includegraphics[width=1.6in]{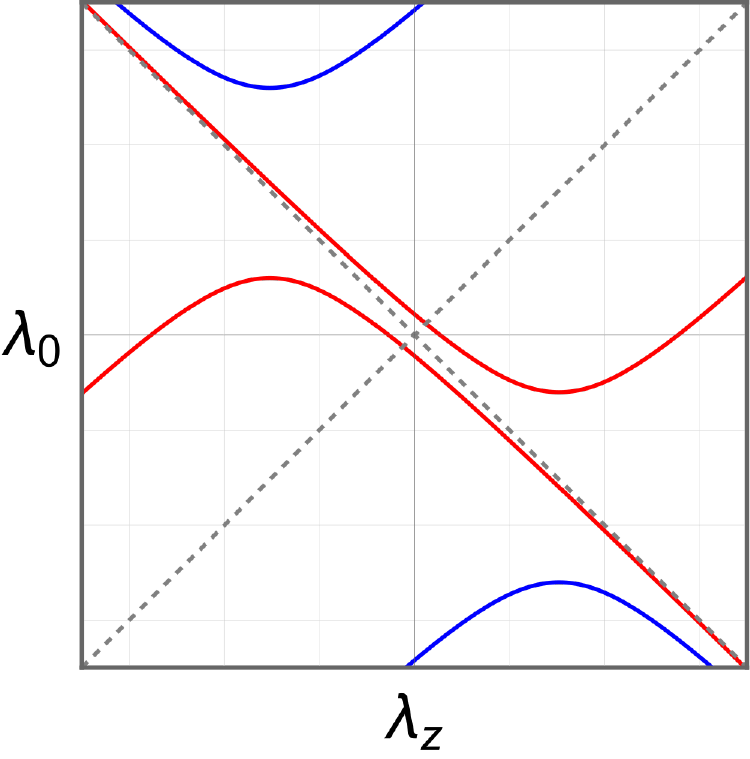}}
\caption{Dispersions for spacelike $b_{\mu}$.}
\label{fig3}
\end{figure}

Next,
suppose that in the original frame 
$b_\mu$ is purely spacelike and perturbative.
Assuming only a nonzero component $b_z$ for simplicity,
the perturbative assumption becomes $|b_z| < m$.
Taking $\lambda_x = \lambda_y = 0$ as before,
the four dispersion branches in relativistic quantum mechanics
are displayed in Fig.~\ref{fig3a}.
In parallel with the perturbative timelike case,
a clean separation is maintained 
between the positive- and negative-energy branches in this frame,
guaranteeing that only positive-energy particles and antiparticles
appear in the field-theoretic description.
Under a sufficiently large observer boost along the $z$ direction,
however,
the magnitude of the components of $|b_\mu|$ grow
and the perturbative condition becomes violated,
producing dispersion branches of the form shown in Fig.~\ref{fig3b}.
The field-theory reinterpretation again generates
particle and antiparticle states of negative energy,
manifesting the concordance problem once more.

The literature contains various suggestions
for avoiding issues associated with negative-energy states. 
A widely adopted and practical choice for the case of finite $m$ 
is to limit attention to scenarios that are perturbative 
in a given initial frame
and to restrict the magnitude of boosts 
so that only frames with perturbative Lorentz violation,
known as concordant frames,
are accessed from the initial frame~\cite{kl01}.
This procedure ensures perturbative consistency essentially by fiat.

Note that in this approach
no mathematical criterion is invoked to select the concordant frames 
from among the infinite set of possible frames.
Instead,
the restriction is physically motivated 
by the apparent lack of large Lorentz violation in nature,
as evidenced by experiments performed in the various observer frames
attained by humanity to date.
These observer frames all involve comparatively small boosts
relative to a canonical Sun-centered frame
\cite{km02},
which itself has comparatively small boost
relative to any cosmic frame associated with the big bang.
It is therefore physically well motivated
to limit attention to concordant frames relative to the Sun-centered frame.
With this additional input,
the restriction to concordant frames 
enables a viable physical description of Lorentz violation.
This approach has permitted numerous measurements 
of coefficients for Lorentz violation to be made 
by experiment and observation within the solar system and beyond
\cite{tables}.

Proposals have also been made to redefine 
the energy surface associated with the vacuum of the free field theory.
In Lorentz-invariant relativistic field theory,
this surface is identified as the zero-energy surface $E=0$,
which contains the state of zero energy-momentum $E= p_j = 0$.
The redefinitions involve
tilting or otherwise changing the shape of the zero-energy surface
so that it becomes dependent on momentum~\cite{dc18,cbj23}.
This approach can ensure a choice of vacuum surface that lies 
below all particle and antiparticle states.

Unlike the approach restricting attention to concordant frames
on physical grounds,
solutions establishing alternative vacuum surfaces
involve judicious mathematical choices
to avoid the appearance of states with negative energies.
One possible advantage is that no physical input is required.
A disadvantage is that the choice of vacuum surface
lacks uniqueness due to the substantial available mathematical freedom. 
Some additional information about this approach 
is provided in Appendix \ref{Existing approaches}.

Note that taking the limit $m\to 0$ in the Lagrange density (\ref{lag})
generates a model that is intrinsically nonperturbative in the above sense.
For instance,
in the field-theory scenario with dispersion relation shown in Fig.~\ref{fig1b},
the negative-energy solutions to the dispersion relation
lie in the vicinity of the apex of the cone.
In this limit,
the concordance problem appears particularly severe
because it is manifest for almost all choices of initial observer frame
and in any boosted frame as well.

\section{Physical resolution}
\label{Physical resolution}

It is evident that any fully satisfactory resolution of the concordance problem
must involve a choice of field-theoretic vacuum
that holds for large coefficients for Lorentz violation,
extends to any observer frame,
and is compatible with the physically relevant realm of interacting fields.
In this section,
we present an alternative resolution satisfying these criteria 
that is inspired by the physical existence of Weyl semimetals. 

\subsection{Weyl semimetals}
\label{Weyl semimetals}

As noted in the introduction,
certain semimetals contain band structures
known to be described in the universal long-wavelength limit
by the dispersion relation~(\ref{disp})
featuring nonperturbative violation of emergent Lorentz invariance.
Since the concordance problem is inherent to this mathematical description,
it is natural to ask how these real physical systems 
evade the corresponding issues.
Here,
we adopt thermodynamic reasoning to address this question. 

In a Weyl semimetal,
the Fermi surface and hence the many-body ground state of the system
are determined thermodynamically 
by the effective heat and particle bath
provided by the semimetal material and its environment,
including the semimetal lattice and excitations
such as phonon and photon modes.
Thermodynamic equilibrium imposes
equality of intensive thermodynamic variables,
such as temperature and chemical potential,
for the electrons and the bath. 
These variables are associated with conserved quantities, 
such as energy and particle number, 
that can be exchanged between the system and the bath.
As a well-known illustration
in the context of the grand canonical ensemble,
the properties of a system having energy levels $E$ and excitation numbers $N$
in equilibrium with a bath of temperature $T$ and chemical potential $\mu$
are governed by the partition function
constructed from factors of the form $\exp[-(E - \mu N)/kT]$,
where $k$ is the Boltzmann constant.

In the ground state of a Weyl semimetal, 
the electrons occupy states with energies up to the Fermi surface.
The energy $E_F$ of the Fermi surface can be identified formally
as the change in the system free energy with excitation number
in the limit $T\to 0$,
or equivalently 
as the value of the chemical potential $\mu$ in the limit $T\to 0$,
$E_F \equiv \mu(T \to 0)$. 
The ground state of a Weyl semimetal
therefore emerges not merely from the pure field theory~(\ref{lag})
of electron excitations 
but also from other fields and interactions forming the thermodynamic bath
via a suitable generalized equilibrium statistical ensemble. 

The correspondence between the fermion number density
and the Fermi surface in a Weyl semimetal
can be obtained by coupling a thermodynamic bath
to the Lorentz-violating theory \eqref{lag}
in the limit $m\to 0$.
We consider here for definiteness
a bath coupling via generic number-conserving interactions. 
The thermodynamic equilibrium state of the fermions is described 
by a density matrix that is diagonal in the fermion energy basis
$\mathinner{|{E_j}\rangle}$.
The occupation probability for the fermion states
is then given by the Fermi-Dirac distribution
\begin{equation}
f(E_j) =
\frac{1}{\mathrm{e}^{(E_j-\mu)/kT}+1} \xrightarrow{T=0} \Theta(E_j-E_F)\,,
\end{equation}
where $\Theta$ is the Heaviside step function.
The chemical potential is obtained implicitly 
by the constraint imposed by the conservation of fermion number $N$ as
\begin{equation}
N = \int D(E) f(E) \mathrm{d} E \xrightarrow{T=0} 
\int^{E_F} D(E) \mathrm{d} E\,,
\end{equation}
where $D(E)= \sum_j \delta(E - E_j)$
is the density of fermion states at energy $E$.

The dispersion equation \eqref{disp} for the fermions
has four solutions, 
which can be written as
\begin{equation}
E_\chi^\pm(\vex{p}) = -\chi b_0 \pm|\vex{p} + \chi \vex b|\,,
\end{equation}
where $\vex p$ is the fermion 3-momentum and
where the chirality $\gamma_5$ has eigenvalues $\chi=\pm 1$
with the signs  $\pm$ labeling the positive and negative branches
of the Weyl cones. 
The density of fermion states for each solution is
\begin{equation}
D_\chi^\pm(E) = V (E + \chi b_0)^2 \Theta(\pm E \pm \chi b_0) /2\pi^2\,,
\end{equation}
where $V$ is the volume of the system.
Summing over all solutions,
we find that the fermion number density $n=N/V$
is related to the Fermi energy $E_F$ 
and to the background fermion density $n_0$ by
\begin{equation}
n=n_0 + \frac1{6\pi^2} \left[(E_F+b_0)^3+(E_F-b_0)^3\right]\,.
\end{equation}
This relation can also be inverted to yield
\begin{equation}
E_F = \sqrt[3]{\tilde n + \sqrt{b_0^6 + \tilde n^2}} + \sqrt[3]{\tilde n - \sqrt{b_0^6 + \tilde n^2}}\,,
\end{equation}
with $\tilde n = 3\pi^2 (n-n_0)/2$.

In the ground state of a Weyl semimetal, 
$n$ and $E_F$ are determined by the underlying chemistry of the material
as well as external potentials and sources of charge.
In the Lorentz-violating field theory \eqref{lag} with $m = 0$,
the requirement of thermodynamic equilibrium with the bath
implies that with $n=n_0$ for $E_F=0$ 
the ground state contains a nonzero density of particles and antiparticles, 
in direct correspondence with the density of electrons and holes
in the semimetal ground state.
In contrast,
in a typical Lorentz-invariant field theory,
the vacuum has no particles or antiparticles
and so $n=n_0=0$, 
which fixes $E_F=0$. 

The above derivations follow standard thermodynamic reasoning,
which implicitly assumes we are working in the rest frame of the bath.
For a semimetal,
choosing this bath frame to determine the Fermi surface and its properties
through thermodynamic arguments is physically appropriate.
This is because
the bath is a fluid of excitations associated with the semimetal itself,
which establishes the bath rest frame as a physically preferred frame.
However,
if instead the bath were somehow independent of the semimetal
and hence uncorrelated with the electronic band structure,
then the mathematics would permit other choices of bath preferred frame,
and the Fermi surface resulting from thermodynamic arguments
would depend on this choice.
We note in passing that this choice 
could be independent of the Lorentz properties of the bath
and could include any available bath frame,
including ones related via conventional Lorentz transformations
invoving the speed of light 
and those related by Lorentz transformations 
in which the speed of light is replaced by the Fermi velocity in the semimetal.

Within the thermodynamic approach advocated here, 
we thus see that
specifying a definite physical ground state of a many-body system
requires two pieces of information.
One involves thermodynamic reasoning,
which establishes that coupling the system to any given bath
determines a ground state. 
The other involves the identification of a physically relevant bath,
which fixes a preferred frame and hence specifies the physical Fermi surface.

\subsection{Fundamental physics}
\label{Fundamental physics}

The Lagrange density (\ref{lag})
provides a mathematical correspondence 
between the descriptions of the semimetal band structure
and the effective field theory for Lorentz violation,
which in turn indicates the existence of a physical resolution 
of the concordance problem in fundamental physics.
The discussion in Sec.~\ref{Weyl semimetals}
then indicates that a thermodynamic definition 
of the physical vacuum in the fundamental theory
would shed light on the relevant issues.

\subsubsection{Establishing the bath}
\label{Establishing the bath}

As a first step,
we replicate the role of the semimetal bath in fixing the Fermi surface 
by incorporating in the fundamental theory a generalized bath
involving one or more additional fields that interact with the fermion $\psi$.
It is conceptually simplest to consider 
the Lorentz-violating model (\ref{lag}) for $\psi$
as being coupled to a Lorentz-invariant bath
having generic Lorentz-invariant interactions with $\psi$
conserving net fermion number,
although in principle the bath and its interactions 
could also incorporate intrinsic Lorentz violation
and could involve more complicated fermion couplings.
For definiteness,
we consider here a generic bath of massless real scalars $\phi$,
so the Lagrange density (\ref{lag}) is extended to
\begin{equation}
\mathcal{L}=
\mathcal{L}_{b}|_{m=0}
+\tfrac{1}{2}\partial_{\mu}\phi\partial^{\mu}\phi
+ \mathcal{L}_{\mathrm{coupling}}\,,
\label{bathlag}
\end{equation}
where $\mathcal{L}_{\mathrm{coupling}}$ 
contains the fermion-bath couplings.
Explicit models with specific interactions 
$\mathcal{L}_{\mathrm{coupling}}$
can be constructed,
an example of which is studied in Appendix~\ref{Explicit model}.

\begin{figure}[t!]
\subfigure[]{\label{fig4a}\includegraphics[width=1.6in]{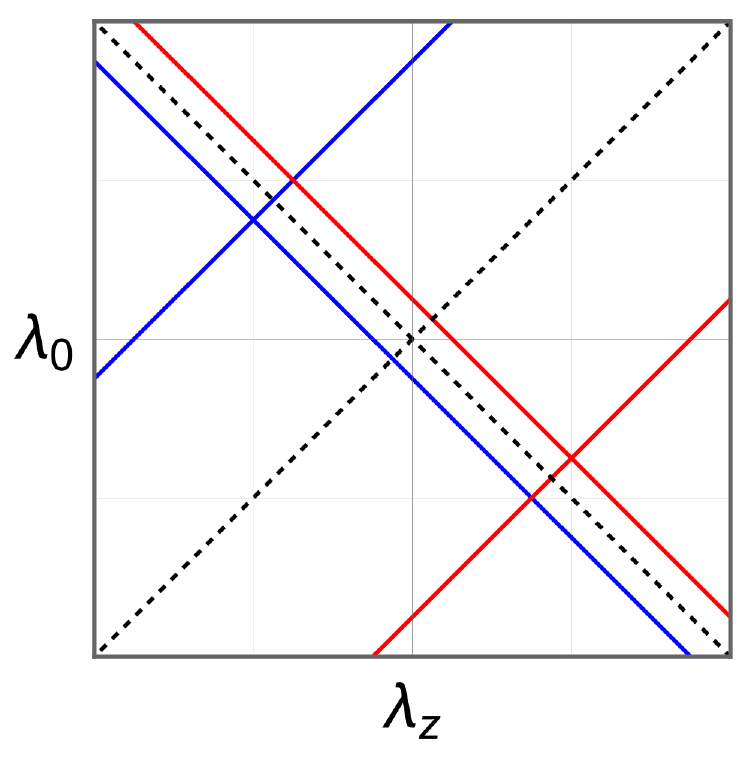}}
\subfigure[]{\label{fig4b}\includegraphics[width=1.6in]{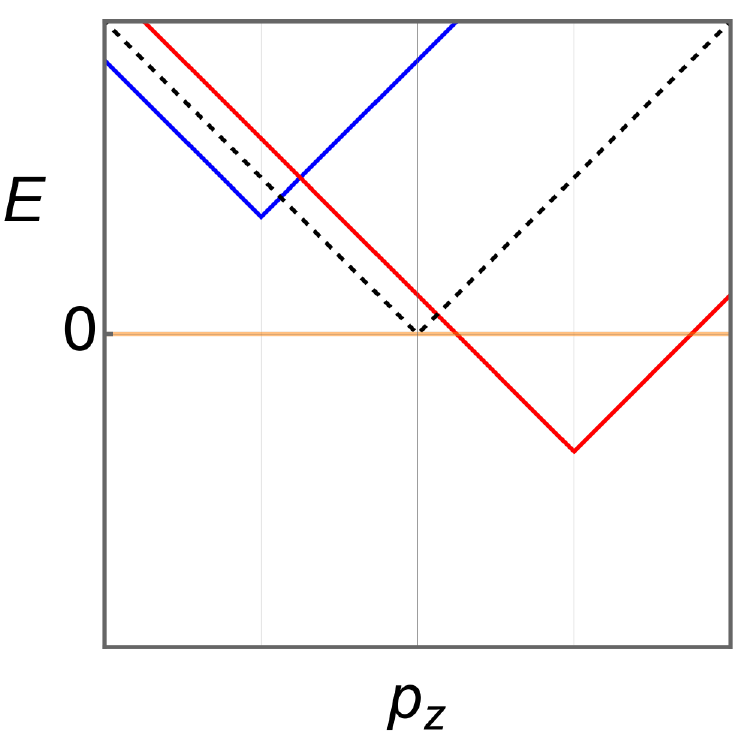}}
\caption{Dispersion relations.}
\label{fig4}
\end{figure}

As a standalone system,
the free scalar theory has dispersion relation along the usual light cone
with vacuum state of zero energy and 3-momentum
$E_\phi = (p_\phi)_j =0$,
along with a conventional vacuum surface $E_\phi = 0$. 
The free-scalar dispersion relation under the conditions 
of Figs.~\ref{fig1a} and~\ref{fig1b}
and for $p_x=p_y=0$ is displayed as dashed lines 
in Figs.~\ref{fig4a} and~\ref{fig4b},
while the fermion dispersion relations are solid lines.
In a generalized statistical ensemble,
the scalar bath has a natural rest frame 
in which the expectation value of the velocity vanishes,
and we adopt this frame for the initial analysis to follow.

When the Lorentz-violating fermion system is coupled 
to this Lorentz-invariant bath,
thermodynamic equilibrium in the generalized statistical ensemble
requires all conjugate variables for the system and bath to match.  
An immediate consequence is that 
the scalar chemical potential is zero for all $T$
because the scalars are real and massless.
The chemical potentials of the fermions and antifermions
must therefore be equal and opposite,
and we can thus take the total fermion chemical potential to vanish as well. 
It follows that the physical vacuum has zero Fermi energy, 
$E_F = 0$, 
which in turn implies that in the physical vacuum
all the negative-energy fermion states must be filled.
In Fig.~\ref{fig4b}, 
these states lie along the two edges of the triangular region
below the zero-energy line.
Note also that if instead $m$ is nonzero and $b_\mu$ is perturbative
with $|b_\mu|\lsim m$ in the chosen frame,
no negative-energy states are present
and the usual perturbative vacuum emerges from the thermodynamic definition,
as expected.

We thus arrive at the conclusion that 
the ground state of the Lorentz-violating quantum field theory~(\ref{lag})
contains physical particles and antiparticles with negative energies,
despite lacking positive-energy particles and antiparticles
as usual.
This result might appear surprising or counterintuitive
in the context of fundamental particle theory,
where the notion of vacuum typically is associated
with the absence of any particle or antiparticle states.
However,
as outlined in Sec.~\ref{Weyl semimetals},
it accurately reflects the physical situation in the Weyl semimetal,
in which the field-theoretic description
reveals occupied physical states lying below the Fermi surface.

With the connection established between the chosen bath and the ground state
of the fundamental theory,
we can next address the question of the selection
of a physically relevant bath.
In the semimetal context,
the appropriate bath rest frame is the rest frame of the semimetal,
as discussed in Sec.~\ref{Weyl semimetals}.
In the fundamental theory,
in contrast,
no direct equivalent exists of the semimetal rest frame. 
The thermodynamic arguments above assume that 
the bath rest frame matches the observer frame in which $b_\mu$ is specified.
However,
no mathematical requirement ensures this match.
Indeed,
in the theory~(\ref{bathlag}),
any bath frame could be chosen from among the infinite set of bath frames
related by Lorentz transformations.
So although the introduction of a definite bath enables
the identification of the ground state of the theory,
additional external input is required to fix the freedom 
of choosing the bath itself.

In a prospective real-world application to fundamental physics,
the role of the bath could be played by a background existing in nature
that is in thermal contact with the system, 
which then fixes the bath rest frame of physical relevance. 
As in the semimetal context,
the bath serves to introduce interactions
that physically establish the ground state of the theory,
thereby avoiding the ambiguities associated with treatments of free fields.
In this context,
a compelling and natural choice for the bath rest frame 
is the known cosmological rest frame,
which coincides with the cosmic microwave background radiation
and is associated with the big-bang fireball.
During the expansion of the Universe,
particles with large Lorentz violation can be expected
to thermalize with other fields and particles in the cosmological fluid,
hence establishing the physically appropriate ground state.
This choice of bath represents a physical proposal
that could in principle be experimentally verified,
should a suitable system with large Lorentz violation
be identified in nature.
Investigation of realistic phenomenological possibilities
to implement these ideas represents an open topic for future work.

The identification of the fireball as the relevant bath
implies an interesting distinction 
between the semimetal and fundamental physics scenarios,
in that the temperature $T$ of a Weyl semimetal
can be varied in the laboratory
while that of the cosmological fluid lies beyond experimental control.
Nonetheless,
in both scenarios the ground state is identified formally 
by taking the limit $T\to 0$ via a generalized statistical ensemble,
irrespective of whether the limit can in fact be approached in nature.

\subsubsection{Restoring concordance}
\label{Restoring concordance}

As discussed in Sec.~\ref{The concordance problem subsection},
one aspect of the concordance problem
concerns the stability of the system with negative energies.
In the present context,
this issue is resolved in the same way 
in both the fundamental particle and semimetal contexts.

We note first that the ground state is stable because 
all negative-energy states are filled.
It follows that no energy can be extracted from the ground state 
by direct transitions from positive- to negative-energy states.

At finite temperature,
some negative-energy states can become vacated.
One might then naively suspect the existence of an instability
in the sense that transitions to empty negative-energy states
would be energetically favored, 
releasing energy to the bath.
However,
stability is unaffected 
because the situation with empty negative-energy states 
represents a system excitation.
As usual for a thermodynamic equilibrium, 
the fluctuation-dissipation theorem~\cite{cw51,rk66}
guarantees that any fluctuations return to the equilibrium state.
For example, 
although at finite temperature a transfer of energy to the bath 
could occur when a particle in a positive-energy state
moves to a vacated negative-energy one,
thermodynamic equilibrium guarantees that this energy is dissipated 
via subsequent excitation from a negative-energy to a positive-energy state. 
Indeed,
stability of the fundamental theory at finite temperature 
is to be expected in light of the close parallel with Weyl semimetals,
which have negative-energy states 
but nonetheless are stable physical systems at finite temperature.

The above perspective reveals 
that the concern about negative-energy states is unfounded.
A system with an energy spectrum unbounded from below
may well have a fatal issue with stability.
However,
the theory (\ref{lag}) has a bounded spectrum
with a well-defined Fock space.
Instead of an issue of stability,
this aspect of the concordance problem is therefore best viewed 
as the challenge of identifying the physical ground state
from among various candidate states in the Fock space.
In conventional Lorentz-invariant quantum field theory,
the vacuum state can be identified through properties
such as its invariance under observer Lorentz transformations.
The latter feature no longer holds 
in the presence of Lorentz violation,
so an alternative approach is needed.
The analogy with Weyl semimetals 
shows that introducing a thermodynamic bath is a useful approach,
as it shifts the ambiguity in selecting a ground state
to the ambiguity in the selection of a bath.
The identification of the physically relevant bath
then resolves the ambiguity,
thereby specifying the ground state
and restoring concordance.

The second aspect of the concordance problem
concerns the effect of large observer boosts.
In a Lorentz-invariant theory,
the size of the observer boost is immaterial
because the vacuum state $E=p_j=0$ is a Lorentz invariant.
However,
a sufficiently large observer boost in a Lorentz-violating theory 
can transform a concordant frame with only positive-energy states
into a nonconcordant one, 
where negative-energy states can appear.
Interpreting the physics of this situation has been deemed problematic,
but the issue can be resolved here 
via the thermodynamic definition of the ground state.

The key point is that the ground state of the fermion system 
is determined by the existence of a bath,
as described in Sec.~\ref{Establishing the bath}.
Once a physically relevant bath has been identified,
the physical fermion vacuum
can be identified through thermodynamic arguments,
which fixes the spectrum of available states.
With the physics established,
observer boosts of any magnitude
amount merely to a change of observer coordinates
and so can at most change the description of the physics,
without changing the physics itself.

As an example,
a suitable boost applied to the situation shown in Fig.~\ref{fig2b}
can result in a description of some negative-energy states 
in the fermion vacuum as positive-energy states in the new frame,
with the physics remaining consistent in both frames.
In this respect the situation has parallels
with other scenarios known to involve the appearance of particles 
when the observer frame is changed. 
For example,
in the Unruh effect 
the vacuum state is perceived by an accelerated observer
as containing a thermal bath of particles,
required for consistent physics between both frames~\cite{wu76,chm07}.
In both these scenarios,
the apparent change in vacuum properties is a consequence
of Lorentz-violating transformations,
one involving intrinsic Lorentz violation
and the other involving acceleration.

Another perspective on boosts is obtained 
by noting that the two types of Lorentz transformations
existing in the absence of the bath,
observer and particle transformations,
can now be supplemented by transformations affecting only the bath.
The bath and particle transformations together
provide the information required to identify the physical fermion vacuum.
The situation in Weyl semimetals is analogous.
In all cases,
the presence of a physically relevant bath ensures that
boosting to the various observer frames results in consistent physics,
thereby resolving the large-boost aspect of the concordance problem.

The effects of a large observer boost can be interpreted
using relativistic statistical mechanics,
with probability factors governed 
by the J\"uttner distribution~\cite{fj11,js57,dc07}.
Following an observer boost to a frame moving at velocity $\vex v$ 
relative to the bath rest frame,
the bath center of mass has 3-velocity $-\vex v$ 
and corresponding 4-velocity $u^\alpha = \gamma(\vex v)(1, -\vex v)$,
with $\gamma^{-1}(\vex v)= \sqrt{1-\vex v\cdot\vex v}$. 
In this boosted frame,
the J\"uttner equilibrium probability factor
$f(E,\vex P, N)$
for a state with $N$ particles can conveniently be written as 
\begin{eqnarray}
f(E,\vex P, N) &=& \exp[-(p_\alpha u^\alpha - \mu_0 N)/kT_0]
\nonumber\\ 
&=& \exp[-(E+\vex v\cdot\vex P-\mu N)/kT]\,,
\label{juttner}
\end{eqnarray}
where $E$ is the total energy 
and $\vex P$ is the total momentum of the system in the boosted frame. 
Here,
$\mu = \mu_0/\gamma(\vex v)$ 
and $T = T_0/\gamma(\vex v)$ 
are related to the
the chemical potential $\mu_0$ 
and temperature $T_0$ of the bath in its rest frame.
Given the occupation number $N_{\vex p} = 0$ or $1$
of the single-fermion state of momentum $\vex p$,
we can write 
\begin{equation}
E = \sum_{\vex p} E_{\vex p}N_{\vex p}\,, 
\quad
\vex P = \sum_{\vex p} \vex p N_{\vex p}\,, 
\quad
N = \sum_{\vex p} N_{\vex p}\,,
\end{equation}
where for simplicity we have disregarded internal degrees of freedom.
The J\"uttner factor~(\ref{juttner})
then separates into a product of factors 
\begin{equation}
f(E,\vex P, N) =
\prod_{\vex p} \exp[-(E_{\vex p} + \vex v \cdot \vex p - \mu) N_{\vex p}/kT]\,. 
\label{fermishift}
\end{equation}
This may be reinterpreted as a change
of the chemical potential from $\mu$ 
to the momentum-dependent value $\mu - \vex v \cdot \vex p$, 
arising from the observer boost.
Note that within this interpretation
the chemical potential $\mu$ and the velocity $\vex v$
remain the appropriate thermodynamical variables in equilibrium.

\begin{figure}
\centering
\subfigure[]{\label{fig5a}\includegraphics[width=1.6in]{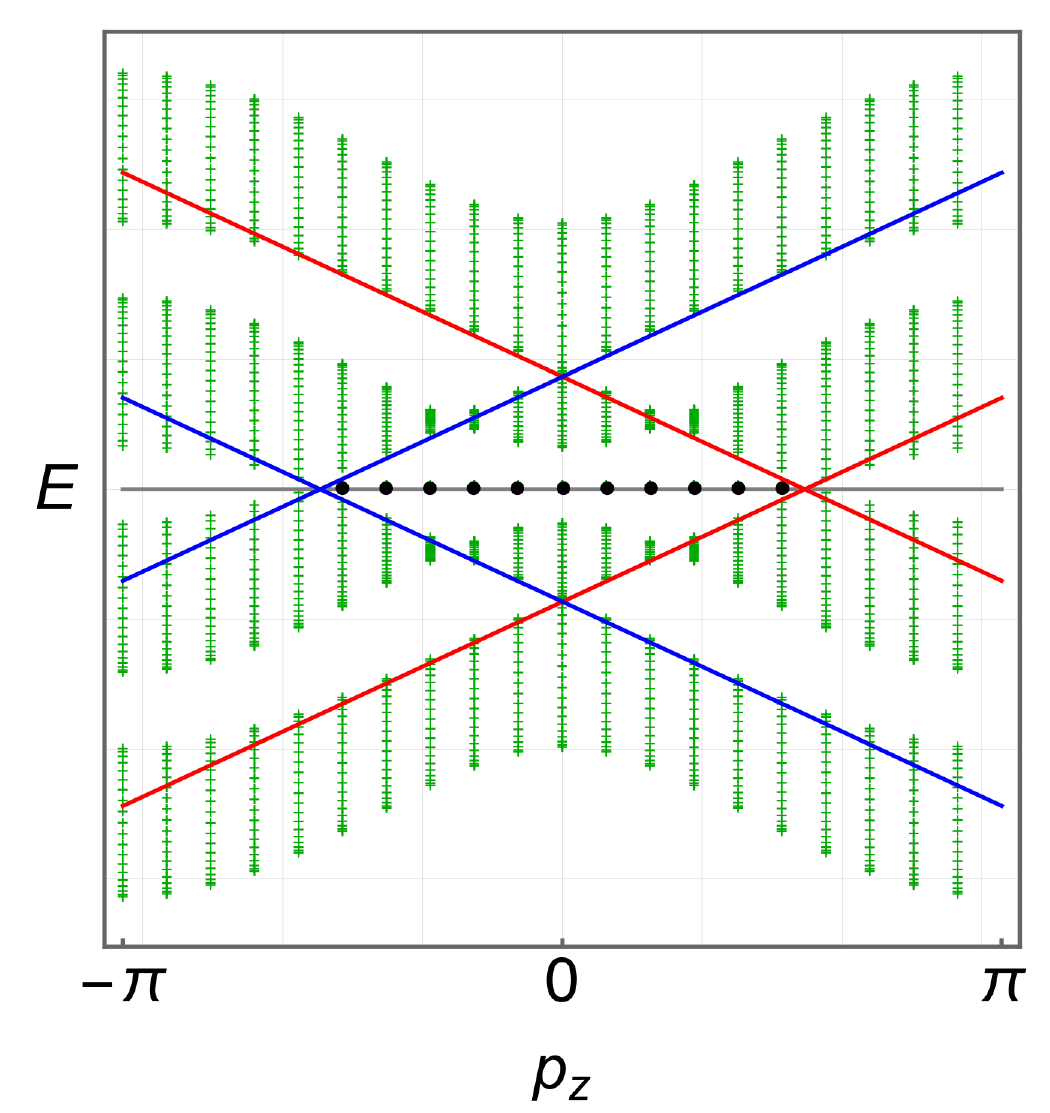}}
\subfigure[]{\label{fig5b}\includegraphics[width=1.6in]{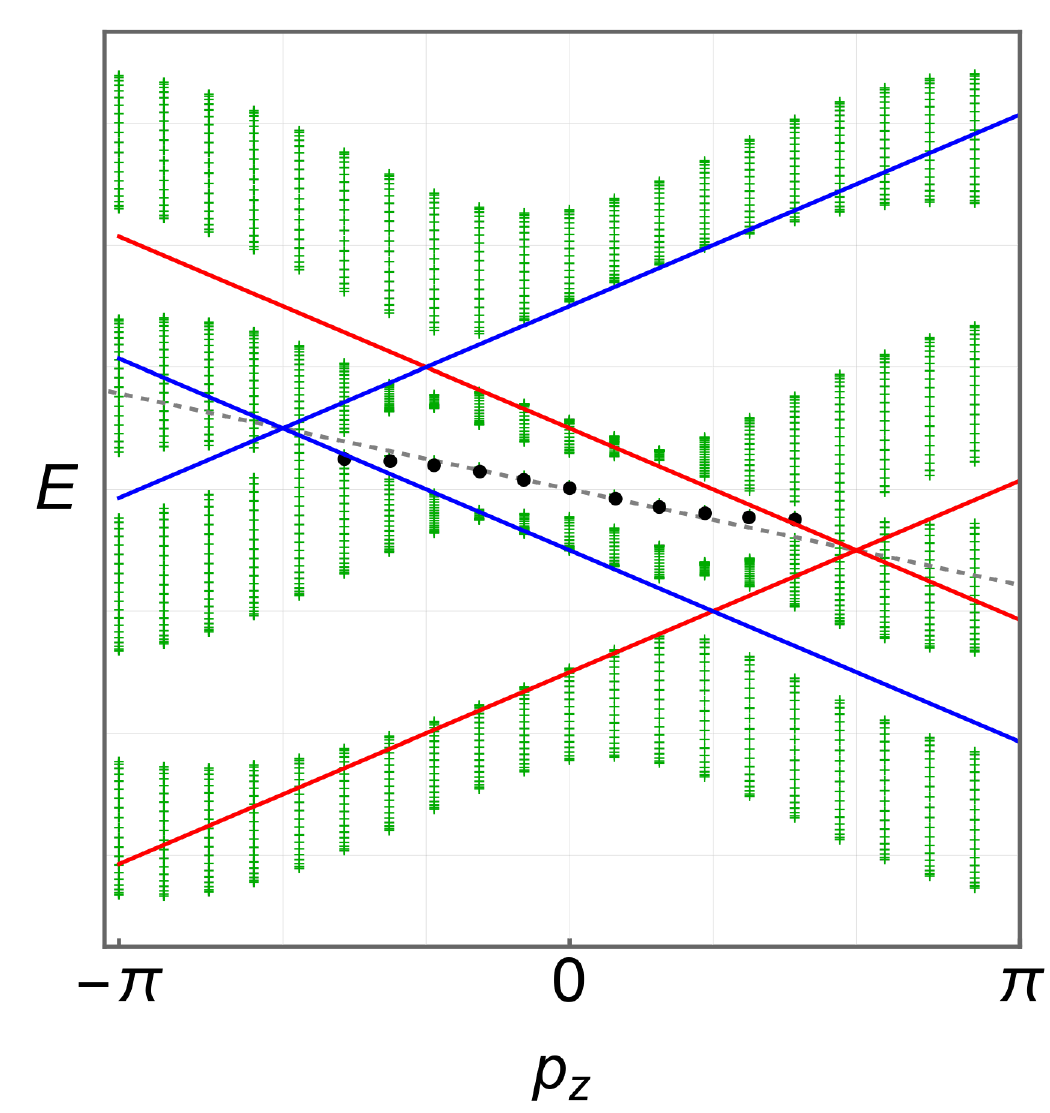}}
\hspace{0.2cm}
\caption{Energy eigenvalues from lattice simulation.}
\label{fig:problem-boosted-weyl-semimetals}
\end{figure}

We can test this description by lattice simulations 
of the hamiltonian of a specific model for a Weyl semimetal~\cite{bskrg19},
which reproduces the Lagrange density $\mathcal{L}_{b}|_{m=0}$
of the theory~(\ref{lag}) in the limit $m\to 0$.
Diagonalizing this hamiltonian numerically 
on a lattice with suitable boundary conditions
leads to a set of bound states on the surface parallel to $\vex b$. 
In the rest frame of the bath, 
these bound states are found to reside at the Fermi level $E_F=0$.
For the spacelike case $b_0 = 0$, $b_z \neq 0$,
the energy eigenvalues in the first Brillouin zone 
are displayed as numerous small crosses in Fig.~\ref{fig5a}, 
while bound states are presented as black dots. 

Performing an observer boost to a frame moving with velocity 
$\vex v = v \hat{\vex z}$ in which $b_0, b_z\neq 0$, 
we find instead the spectrum shown 
in Fig.~\ref{fig5b}.
The bound states no longer reside at $E_F=0$, 
but instead span the dashed line connecting the Weyl nodes.
Thus, 
the occupations of the states of the Weyl semimetal 
in the boosted frame can be inferred from those in the bath rest frame, 
with the reinterpretation of the Fermi level as $E_F- \vex v \cdot \vex p$,
in agreement with the result~(\ref{fermishift}).
This confirms that the ground state itself is observer independent, 
even though its coordinate description changes in the boosted frame.

Note also that some bulk positive-energy states in the bath rest frame 
become negative-energy states in the boosted frame. 
However,
the reinterpreted Fermi level ensures that these states remain unoccupied
in the boosted frame.
This provides another verification
that the descriptions in the two frames are physically equivalent.

\section{Implications}
\label{Implications}

The above treatment implies several interesting features 
of the physical ground state in the presence of large Lorentz violation.
In this section,
we offer some comments on generic properties,
on topological properties,
and on prospective physical implications,
primarily in the context of the theory (\ref{lag}).

\subsection{Generic properties of the ground state}
\label{General properties of the vacuum}

By virtue of the global U(1) phase invariance 
under the transformation 
$\psi\to \exp(\mathrm{i}\alpha)\psi$ for real $\alpha$,
the theory (\ref{lag}) exhibits a conserved current
\begin{equation}
j^{\mu}=\bar{\psi}\gamma^\mu\psi\\, \qquad \partial_\mu j^{\mu}=0\,.
\end{equation}
The corresponding conserved charge is the fermion number.
Provided the coupling to the bath is also U(1) invariant,
the system-bath combination governed by the Lagrange density~(\ref{bathlag})
also conserves fermion number.
Note that scalar particle number is not conserved in this model
because the scalars are real. 
At finite temperature $T$, 
the system-bath combination 
contains fermion-antifermion pairs and scalars in equilibrium,
and in the limit $T\to 0$ only negative-energy pairs remain in the vacuum. 
These considerations reveal that the number densities 
of particles and antiparticles are equal in the ground state. 

Another feature of the ground state is that
the net velocity of the particles and antiparticles is zero.  
To see this,
recall again that thermodynamic variables 
for a generalized statistical ensemble
are associated with overall conserved quantities 
that can be exchanged between system and bath.
For example,
the temperature arises
from the exchange of conserved energy.
In the present instance,
the Lagrange density~(\ref{bathlag})
for the Lorentz-violating fermion system and the bath
involves interactions invariant under spatial translations,
so 3-momentum is exchanged and conserved.
In a generalized statistical ensemble,
the thermodynamic variable associated with
exchanging the conserved 3-momentum $\mathbf{p}$ with the bath
is the group 3-velocity $\mathbf{v}$,
which can be defined as the change of free energy with $\mathbf{p}$. 
This can also be seen from the relativistic grand canonical ensemble,
which is governed by the J\"uttner distribution~(\ref{juttner})
containing factors of the form $\exp[-\vex p \cdot \vex v/kT]$.
It follows that in thermodynamic equilibrium 
the scalar and fermion velocities must be equal, 
$\mathbf{v}_\phi=\mathbf{v}_\psi$.
Since the Lorentz invariance of the scalar ensures 
that $\mathbf{v}_\phi$ is zero in the vacuum,
the total fermion 3-velocity must vanish in the vacuum as well.

Note that the vanishing of the total fermion 3-velocity 
is a consequence of thermodynamic equilibrium
and the choice of physical bath.
It holds even though the total fermion momentum 
in the ground state is nonzero.
The nonvanishing of the total fermion momentum
can be seen directly from Fig.~\ref{fig4b},
as the cone containing the negative-energy states
lies off axis in 3-momentum space. 
This discrepancy between velocity and momentum
is a well-established generic aspect of Lorentz-violating theories.
It can be traced to the associated pseudo-Finsler geometry
that replaces the conventional Minkowski geometry 
of Lorentz-invariant theories~\cite{ak04,ak11,ac05,kr10}. 

Note also that the above features
hold even when the fermion carries a charge 
through an additional coupling to a gauge field.
This demomstrates that the ground state 
in the presence of large Lorentz violation
remains a state of zero total charge and zero total current,
in parallel with conventional Lorentz-invariant scenarios.

\subsection{Topological properties of the ground state}
\label{Topological properties of the vacuum}

The presence of the Weyl nodes arising from large Lorentz violation
endows the ground state of the theory~(\ref{lag}) 
with certain geometric and topological properties
that lead to potentially observable effects.
One qualitatively distinct feature 
is the existence of a nonzero Pancharatnam-Berry 
geometric phase~\cite{sp56,mb84,xcn10}
associated with the physical vacuum.
In general,
for a given closed path $\mathcal{C}$ at fixed energy in 3-momentum space,
the geometric phase $\Phi$ can be defined 
via the line integral of the expectation value 
of the position operator,
\begin{equation}
\Phi=\oint_{\mathcal{C}} \mathrm{d}\mathbf{l} \cdot \langle
\psi(\mathbf{p})|\mathrm{i} \nabla_{\mathbf{p}}| \psi(\mathbf{p})
\rangle \,.
\end{equation}
A fermion with momentum evolving around the path $\mathcal{C}$
accumulates a net phase change $\Phi$,
which in principle is experimentally observable via interferometry.

As an explicit example,
in the theory~(\ref{lag})
$\Phi$ is nonzero whenever a fermion with $p_z=0$ adiabatically traverses 
a closed path in momentum space 
about the special momentum $\mathbf{p}_{\rm s}=\mathbf{b}$
fixed by the minimum energy $E_{\rm s}=-b_0$.
The point $(E_{\rm s}, \mathbf{p}_{\rm s})$
is located at the apex of the cone of negative-energy states,
and the closed path lies on the momentum 2-sphere
displayed as a circle in Fig.~\ref{fig1b}.
Calculation shows that 
the corresponding geometric phase for a fermion 
can be taken as $\Phi =\pi$ 
when $\mathcal{C}$ encloses $\mathbf{p}_{\rm s}$
and $\Phi = 0$ otherwise~\cite{klss24-2}.

We can therefore conclude that the definition of the physical ground state 
through the interaction of the fermion system with the bath 
incorporates geometric phases
for all negative-energy particles and antiparticles.
It follows that the physical ground state is uniquely identified 
up to an overall phase
only when the corresponding relative geometric phases are specified.

Note also that a fermion acquires an analogous nonzero phase of opposite sign
if instead $\mathcal{C}$ encloses
the apex of the positive-energy cone displayed in Fig.~\ref{fig1b} 
The appearance of an observable geometric phase can therefore be viewed
as originating from the separation of the Weyl nodes 
due to large Lorentz violation.

\subsection{Physical implications}
\label{Physical implications}

The presence of negative-energy fermions and antifermions in the vacuum 
raises the question of their prospective experimental observability.
Although the theory~(\ref{lag}) is a model constructed primarily
to illustrate the theoretical resolution of the concordance problem
using physical reasoning,
observable effects could in principle appear
in a more realistic theory of fundamental particles.
In practice,
however,
such effects may be challenging to detect in the laboratory.

Consider the theory~(\ref{lag}) in the limit $m\to 0$
and suppose,
for example,
that the fermion $\psi$ carries electric charge.
Applying an electromagnetic field $F_{\mu\nu}$ to the ground state 
then generates a 4-current $j^\mu$ 
of negative-energy fermions~\cite{kl16,bskrg19},
$j^{\mu} = 
{\alpha}
\varepsilon^{\mu\nu\rho\sigma} b_{\nu}F_{\rho\sigma}/{\pi}$, 
where $\alpha$ is the fine-structure constant,
along with a corresponding antifermion current.
In terms of the electric field $\vex E$ and the magnetic field $\vex B$, 
this gives
\begin{equation}
j^0 = \frac{2\alpha}{\pi}\mathbf{b}\cdot\mathbf{B}\,,
\qquad
\mathbf{j} = \frac{2\alpha}{\pi}(b_0\mathbf{B}-\mathbf{b}\times\mathbf{E})\,.
\end{equation}
A magnetic field thus produces a current 
parallel to $\vex B$ if $b_0$ is nonvanishing,
while an electric field with a component orthogonal
to a nonzero $\vex b$ generates a current transverse to $\vex E$.

Chiral electromagnetic effects of these types
might in principle be measurable.
As a rough approximation to a maximal attainable sensitivity,
suppose an experiment involves electric field strengths
of order $|\vex E| \simeq 10$ V/m 
in a focused beam area $A \simeq (\unit[1]{\upmu m})^2$,
comparable to those attainable 
in a powerful optical laser~\cite{rj21}
or X-ray free-electron laser~\cite{glhp22},
or magnetic field strengths of order $|\vex B| \sim$ 100 T
as achieved with pulsed-field magnets~\cite{nhmfl}.
Taking currents of order $I \simeq \unit[10^{-10}]{A}$ 
as being experimentally measurable
then yields crude maximal attainable sensitivities
to the components of $b_\mu$ 
of order $|b_0| \approx |\vex b| \simeq 10^{-20}$ GeV.
These values for large Lorentz violation
exceed by about a billionfold
the best existing precision measurements of $b_\mu$ components
for perturbative Lorentz violation 
in some charged-particle systems~\cite{tables}.
It is nonetheless evident that detecting
effects from large Lorentz violation 
in the laboratory can be expected to be challenging.

The fermion $\psi$ in the theory~(\ref{lag}) 
cannot be identified with the electron field in a laboratory setting 
because the electron mass $m_e \simeq 511$ keV
exceeds by over 20 orders of magnitude the existing constraints~\cite{tables}
on components of the electron coefficient $b_\mu$,
so an observer boost with gamma factor surpassing $10^{20}$
would be required to achieve a nonconcordant frame
with negative-energy states in the vacuum,
and furthermore these would be populated
only if the bath were similarly boosted as well.
Although electrons can be effectively massless under extreme circumstances 
such as in the early Universe above the electroweak phase transition,
the components of $b_\mu$ would need to exceed the fireball temperature
to have prospects of significant effects. 
For similar reasons, 
$\psi$ cannot represent any known charged lepton or baryon field.
Even assuming the existence of a hypothetical massless fermion
with large Lorentz violation generated by a coefficient $b_\mu$ 
that is tiny compared to known mass scales,
the corresponding occupied negative-energy states
would be difficult to detect in the laboratory 
because they represent a spectrum of ultrasoft massless particles.
On astrophysical or cosmological scales,
however,
and in a version of the theory~(\ref{lag})
suitable for light neutral particles such as neutrinos,
the negative-energy states might help resolve open issues
such as the tension between laboratory and astrophysical measurements
of neutrino masses or the nature of dark energy. 

In any case,
the above considerations illustrate the challenge 
of using direct laboratory measurements
to verify the existence of negative-energy states 
in the physical vacuum of a fundamental particle theory.
This situation can be contrasted,
for example,
with typical Weyl semimetal phases~\cite{lqd21},
which in units with $c=1$ have timelike component $|b_0|$ of order 100 meV 
and spacelike components $|\mathbf{b}|$ of order 100 eV
that are perturbative with respect to $m_e$
but far exceed the zero effective mass
of excitations near the Weyl nodes.
In this respect,
Weyl semimetals represent a unique environment 
within which to study experimentally 
the pivotal conceptual issues
associated with large Lorentz violation.

\section{Summary}
\label{Summary}

The development over the last few decades
of a physically realistic effective field theory
to describe general perturbative Lorentz violation in fundamental physics
has inspired many new experimental searches
for the corresponding effects~\cite{tables}.
This success has in turn brought to the forefront the concordance problem:
how physically to interpret models with nonperturbative Lorentz violation,
which involves scenarios where the corresponding coefficients
either are large in a given observer frame
or become large in a highly boosted frame.

In this work,
we have addressed the concordance problem from a physical perspective,
taking advantage of the mathematical equivalence 
between fundamental particle models with nonperturbative Lorentz violation
and treatments of certain Weyl semimetals in the condensed-matter context.
In the latter,
Lorentz invariance is an emergent feature
of the electronic band structure,
and nonperturbative deviations from exact symmetry
are readily achieved in the laboratory.

Our focus here is the theory~(\ref{lag}),
which has been widely adopted to describe the effects of Lorentz violation
in both the condensed-matter and the fundamental contexts.
The manifestations of the concordance problem in this theory
are characterized.
We then determine its resolution in Weyl semimetals 
using equilibrium thermodynamics.
The insights we acquire from this treatment
provide a guide to a physical resolution of the concordance problem
in fundamental physics.
The key point is the introduction of a thermodynamic bath,
which enables transferring issues with the theory
to properties of the bath. 
This idea by itself entails a dependence of the resulting ground state
on various bath properties.
Physical applications typically incorporate the relevant bath,
such as the cosmological fluid from the big bang,
which permits a definitive identification of the ground state
and hence a resolution of the issues. 
We discuss some implications of this treatment,
including generic, topological, and physical aspects
of the ground state.

In conclusion,
we have shown here that the existence of Weyl semimetals
and considerations of thermodynamic equilibrium 
offer a physical resolution for the concordance problem 
associated with stability and large boosts
in a widely studied Lorentz-violating theory of fermions.
We find it remarkable and beautiful that the key 
to a consistent physical interpretation 
of large explicit Lorentz violation in a fundamental theory
resides in broad thermodynamic considerations 
and can ultimately be traced 
to the physical breaking of global Lorentz invariance
caused by the cosmological background.
In any event,
the prospects appear excellent that future work along related lines
will provide a complete resolution of the concordance problem
for other types of Lorentz violation in fundamental theories.

\begin{acknowledgments}
This work is supported in part 
by the U.S.\ Department of Energy 
under grant {DE}-SC0010120,
by grants FAPEMA Universal 00830/19 
and CNPq Produtividade 310076/2021-8,
by CAPES/Finance Code 001,
and by the Indiana University Center for Spacetime Symmetries,
College of Arts and Sciences, 
and Institute for Advanced Study.
\end{acknowledgments}

\appendix

\section{Aspects of concordance}
\label{Existing approaches}

In this appendix,
we first provide an example of a typical issue 
associated with large observer boosts.
An outline is then given of an existing formal approach
to the choice of vacuum.

Consider for definiteness the kinematics of the unconventional process
involving a fermion $f$ radiating a fermion-antifermion pair, 
$f \to f f \overline{f}$.
We suppose that $f$ is a particle created by the fermion field $\psi$
in the theory~(\ref{lag})
and that a suitable 4-point interaction enabling the fermion breakup is present.
This process can be kinematically allowed
in the presence of Lorentz violation.
It suffices for our purposes to consider a collinear scenario
where the particles move along a single spatial direction.

Let $k$ be the momentum of the incoming fermion
and $q_1$, $q_2$ the momenta of the radiated fermion and antifermion,
respectively.
The energy-momentum state of the antifermion 
is assumed to sit on the same dispersion branch as those of the fermions. 
Momentum conservation dictates 
that the final momentum of the radiating fermion is $k-q_1-q_2$.
Defining the energy difference $\Delta E$ as 
\begin{eqnarray}
\Delta E &=& E(k)-E(q_1)-E(q_2)-E(k-q_1-q_2)\,,
\nonumber\\[1ex]
E(k) &=& \sqrt{(b_0-k)^2+m^2}-b_z\,,
\label{eq:energy-balance}
\end{eqnarray}
we see that fermion breakup is kinematically possible 
when configurations $\{k,q_1,q_2\}$ exist 
that satisfy the conservation of energy,
$\Delta E=0$. 
For example,
a possible configuration kinematically allowing fermion breakup is~\cite{kl01}
$k= 3q_1 = 3q_2 = 2[m^2+(b_0)^2]/b_0$.

\begin{figure}
\centering
\subfigure[]{\label{figa1a}\includegraphics[width=1.6in]{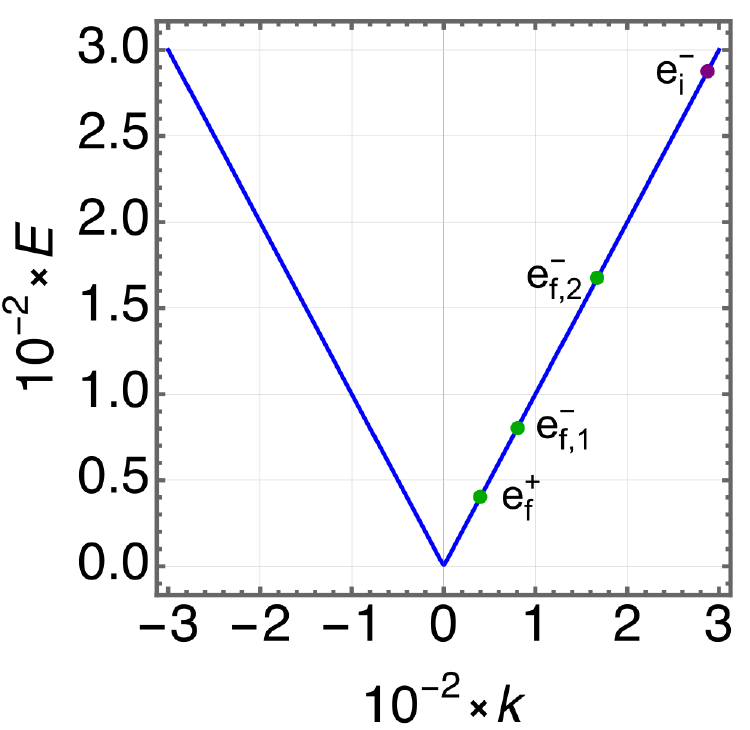}}
\subfigure[]{\label{figa1b}\includegraphics[width=1.6in]{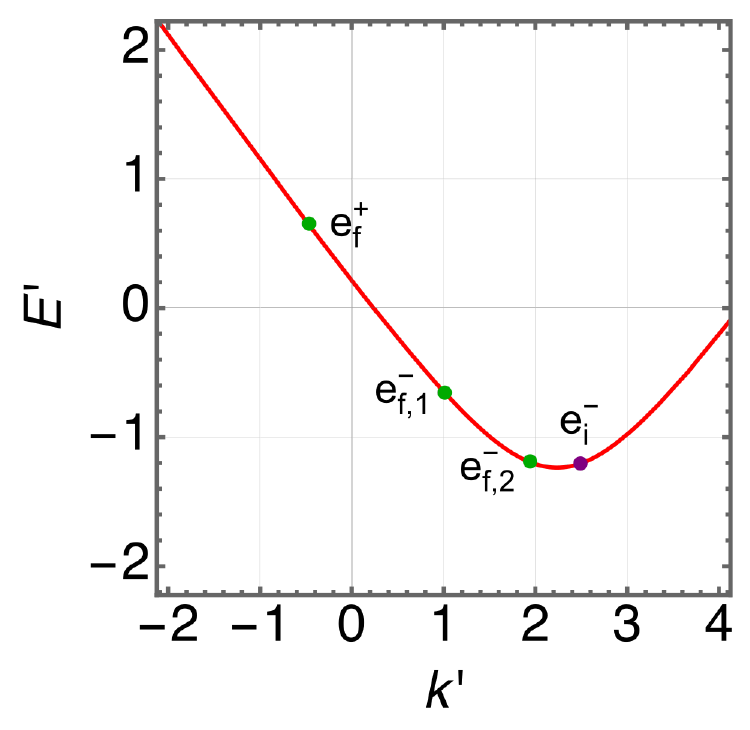}}
\caption{Fermion and antifermion energies and momenta.}
\end{figure}

For illustrative purposes,
we study first the regime of perturbative Lorentz violation
in a concordant frame,
with $b_0=10^{-2} m$, $b_z=0$. 
Working with numerical values for energy and momentum
relative to the fermion mass $m$,
one configuration with all spatial momenta different
that satisfies energy conservation $\Delta E=0$ is
$k\simeq 287.31m$, $E(k)\simeq 287.30m$,
$q_1=80m$, $E(q_1)\simeq 80.00m$,
$q_2=40m$, $E(q_2)\simeq 40.00m$,
$k-q_1-q_2\simeq167.31m$, and $E(k-q_1-q_2)\simeq 167.30m$.
Note that all the energies are positive
and that all energies and momenta are far greater than $m$.
This scenario is illustrated in Fig.~\ref{figa1a}.
The fermion and antifermion energies and momenta appear 
as dots lying far from the dispersion minimum,
which on this scale appears as a cusp at the origin
despite the presence of a nonzero fermion mass.

Implementing a sufficiently large observer boost from a concordant frame
to a nonconcordant one achieves a regime of large Lorentz violation.
For example,
we can perform a boost with $\beta_b=0.99999$ 
such that the transformed components of $b_\mu$ are
$b_0'=\gamma_b(\beta_b)b_0\approx 2.24m$,
$b_z'=\beta_b\gamma_b(\beta_b)b_0\approx 2.24m$.
These values are larger than the fermion mass,
so Lorentz violation is large in this frame.
Applying this observer boost to the energies and momenta
in the original concordant frame gives
$k'\simeq 2.49m$, $E'(k')\simeq -1.20m$,
$q_1'\simeq 1.02m$, $E'(q_1')\simeq -0.66m$,
$q_2'\simeq -0.47m$, $E'(q_2')\simeq 0.65m$,
$k'-q_1'-q_2'\simeq 1.94m$, and $E'(k'-q_1'-q_2')\simeq -1.19m$.
We see that the norms of the energies and momenta
are observed to be less than or of order $m$ in the nonconcordant frame.
This boosted scenario is shown in Fig.~\ref{figa1b}. 
In contrast to the situation in Fig.~\ref{figa1a}, 
the nonconcordant dispersion plotted in Fig.~\ref{figa1b} 
includes negative values for the energy.
Indeed,
three of the four energies displayed as dots
become negative in this frame. 
The treatment of the corresponding states
as positive-energy particle states is thus obscured, 
which in turn raises issues about the physics of fermion breakup in this frame. 
This example illustrates the boost aspect of the concordance problem
introduced in Sec.~\ref{The concordance problem subsection},
which amounts to the challenge 
of providing a consistent physical interpretation
across concordant and nonconcordant frames.

An ingenious formal approach to resolving the concordance problem
that relies on concepts of classical hamiltonian dynamics
has been presented by Colladay~\cite{dc18}.
In a Lorentz-invariant theory, 
the classical canonical hamiltonian $H^{\rm c}$ 
of a relativistic point particle vanishes identically, 
$H^{\rm c}\equiv 0$, 
due to reparameterization invariance of the particle worldline.
Instead,
an extended hamiltonian $H^{\rm ext}$ 
can be constructed following Dirac~\cite{pd50,hrt76},
in which the dispersion equation is understood
as a primary first-class constraint.

This chain of reasoning can be applied to scenarios with Lorentz violation, 
as the corresponding classical-particle actions 
are still invariant under reparametrizations of the worldline~\cite{kr10}.
For finite $m$,
the dispersion relation for the theory~(\ref{lag})
decomposes into two observer-covariant factors~\cite{dc18}
according to
\begin{eqnarray}
&R_+(\lambda) R_-(\lambda) = 0\,,
\nonumber \\
&R_{\pm}(\lambda) =
\tfrac{1}{2}\left(\lambda^2-m^2
-b^2\pm 2\sqrt{(b\cdot \lambda)^2-b^2\lambda^2}\right)\,.
\qquad
\end{eqnarray}
Two extended hamiltonians can then be constructed, as
\begin{equation}
\mathcal{H}_{\pm}^{\mathrm{ext}}=-\frac{e}{m}R_{\mp}(\lambda,x)\,,
\end{equation}
where $e$ is an auxiliary function
having specific form inessential to our purpose.
Note that the two hamiltonians $\mathcal{H}_{\pm}^{\mathrm{ext}}$
vanish on shell but are nonzero off shell,
so they are useful off-shell quantities. 

\begin{figure}
\centering
\subfigure[]{\label{figa2a}\includegraphics[width=1.6in]{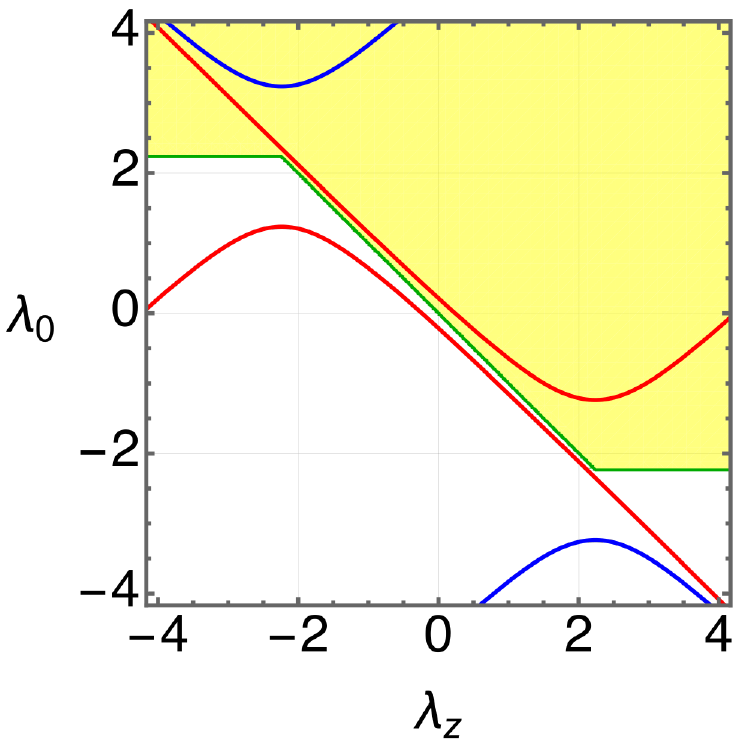}}
\subfigure[]{\label{figa2b}\includegraphics[width=1.6in]{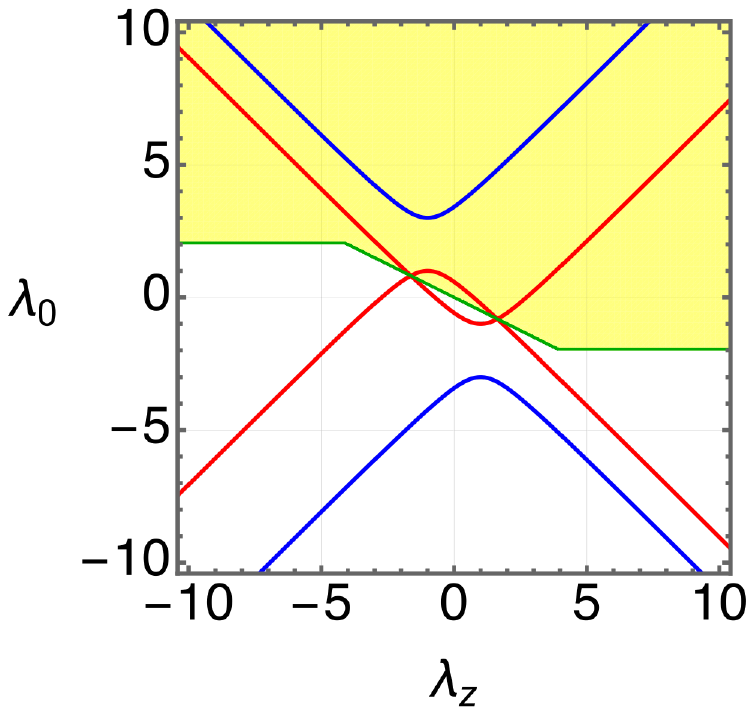}}
\caption{Off-shell positivity regions.}
\end{figure}

In a concordant frame,
one of the two extended hamiltonians 
describes the positive-energy dispersions
while the other describes the negative-energy ones.
Since the two factors are observer independent,
the two hamiltonians can be used
to separate positive- and negative-energy branches in any frame.
To establish this,
we can define
\begin{equation}
\Lambda_{\pm}(\lambda)
:=-\frac{2m}{e}\frac{\partial\mathcal{H}_{\pm}^{\mathrm{ext}}}
{\partial \lambda^0}\,.
\end{equation}
Using one of the Hamilton covariant equations of motion,
$u_{\pm}^{\mu}=
-{\partial\mathcal{H}^{\mathrm{ext}}_{\pm}}/{\partial \lambda_{\mu}}$,
allows us to deduce that
\begin{equation}
\Lambda_{\pm}(\lambda)|_{\text{on-shell}}>0\,.
\label{eq:positivity-region}
\end{equation}
This statement is powerful because 
it holds independently of the observer frame. 
Indeed,
the condition $\Lambda_{\pm}(\lambda)>0$ 
fixes an off-shell positivity region in momentum space
that contains the on-shell positive-energy dispersions 
in an observer-invariant way,
as demonstrated by Eq.~\eqref{eq:positivity-region}.
As a result,
the boundary of the positivity region
can be adopted by fiat as an observer-independent choice
for the energy surface of the vacuum of the theory,
even in the presence of large Lorentz violation.
 
For many values of $b_\mu$,
the boundary of the positivity region
cleanly separates the positive- and negative-energy states.
This situation is illustrated in Fig.~\ref{figa2a}
for large Lorentz violation in a chosen frame.
The conventional choice of vacuum surface in this frame
would lie at $E=0$
and would thereby incorporate both positive- and negative-energy states
in two of the four dispersion branches.
Note,
however,
that if both the individual components $b_{\mu}$ 
and the magnitude $b^2$ are large in a given frame, 
the positive- and negative-energy dispersions 
can intersect.  
This situation is shown in Fig.~\ref{figa2b}.
The above formal procedure then still provides
an observer-independent separation,
but a satisfactory dispersion with positive energies
may require contributions from different branches 
such that the resulting curve lies completely in the positivity region. 
This introduces points where the dispersion curve is nondifferentiable,
which may be physically problematic.

\section{Explicit model}
\label{Explicit model}

This appendix considers an explicit model
with a particular interaction between the fermion and the scalar bath.
We demonstrate that a generic initial configuration
of fermion and antifermion states
can indeed reach the thermodynamic vacuum as the system is cooled,
while conserving 4-momentum in all interactions.

A comparatively simple choice for the fermion-scalar coupling 
$\mathcal{L}_{\mathrm{coupling}}$ 
in Eq.~(\ref{bathlag}) is the Yukawa interaction,
which is Lorentz invariant and conserves net fermion number.
We adopt here the Lagrange density
\begin{equation}
\mathcal{L}_{\mathrm{model}}=
\mathcal{L}_{b}|_{m=0}
+\tfrac{1}{2}\partial_{\mu}\phi\partial^{\mu}\phi+g\overline{\psi}\psi\phi\,.
\label{lag2}
\end{equation}
Quantization of this model reveals that the particle spectrum
includes fermions $f$ and antifermions $\overline{f}$,
along with scalar particles $\sss$ representing excitations of the bath.
The states of the system-bath combination 
lie along the dispersion relations shown in Fig.~\ref{fig2b}.
According to the thermodynamic definition of the physical vacuum,
all the fermion and antifermion states with negative energies are occupied
in the ground state,
while those with positive energy are empty. 
At finite temperature,
the fermions and antifermions are instead distributed among all the states 
according to a generalized statistical distribution. 

\begin{figure}
\subfigure[]{\label{figb1a}\includegraphics[width=0.8in]{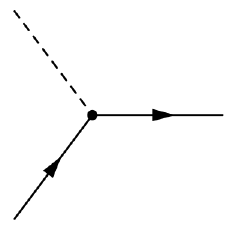}}
\subfigure[]{\label{figb1b}\includegraphics[width=0.8in]{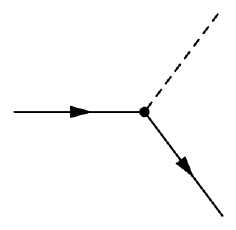}}
\subfigure[]{\label{figb1c}\includegraphics[width=0.8in]{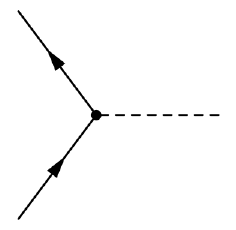}}
\subfigure[]{\label{figb1d}\includegraphics[width=0.8in]{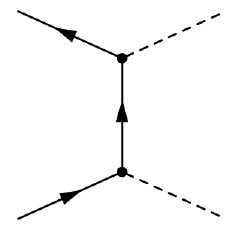}}
\caption{Fermion-scalar processes.}
\end{figure}

Three basic kinds of processes involving $f$, $\overline{f}$, and $\sss$
arise from the Yukawa interaction in Eq.~\eqref{lag2}.
First, the emission of scalars via the process $f\rightarrow f+\sss$ 
as in Fig.~\ref{figb1a}
represents a loss of fermionic energy and momentum.
Second, the absorption of scalars via $f+\sss \rightarrow f$
as in Fig.~\ref{figb1b}
represents a gain of fermionic energy and momentum.
Finally, 
the pair annihilation $f\overline{f}\rightarrow n\sss$ 
to $n$ scalars can occur 
via tree-level Feynman diagrams with $n$ interaction vertices.
The cases with $n=1$ and~2 are shown
in Figs.~\ref{figb1c} and~\ref{figb1d}.
These processes are constrained by the requirement
of conservation of 4-momentum among the particles involved.

By construction,
the physical ground state has zero net fermion number.
The issue of whether this state can be attained
as the temperature is reduced
therefore amounts to determining whether the allowed processes suffice
to ensure that any $f$ or $\overline{f}$ with positive energy 
can either settle in a negative-energy state or annihilate.

\begin{figure}
\centering
\subfigure[]{\label{figb2a}\includegraphics[width=2.2in]{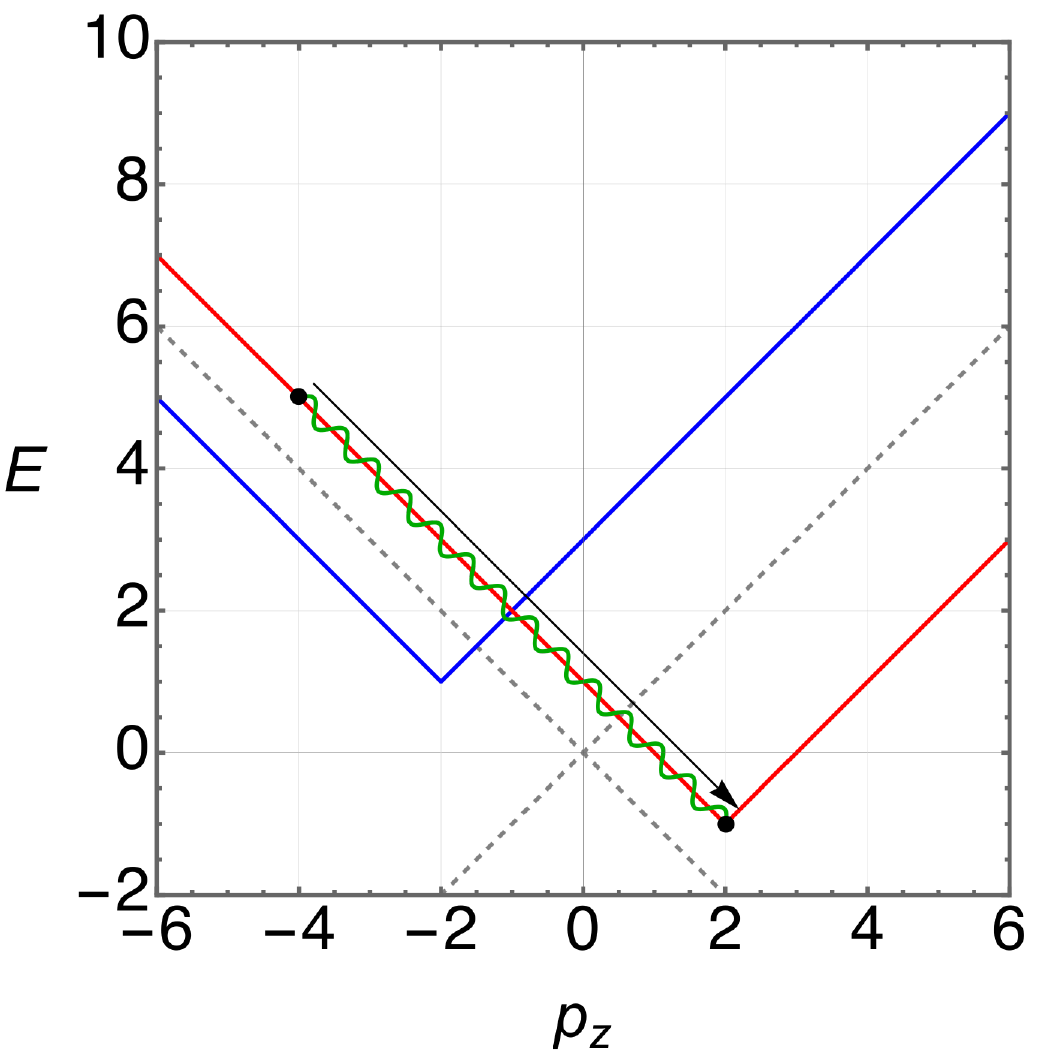}}
\subfigure[]{\label{figb2b}\includegraphics[width=2.2in]{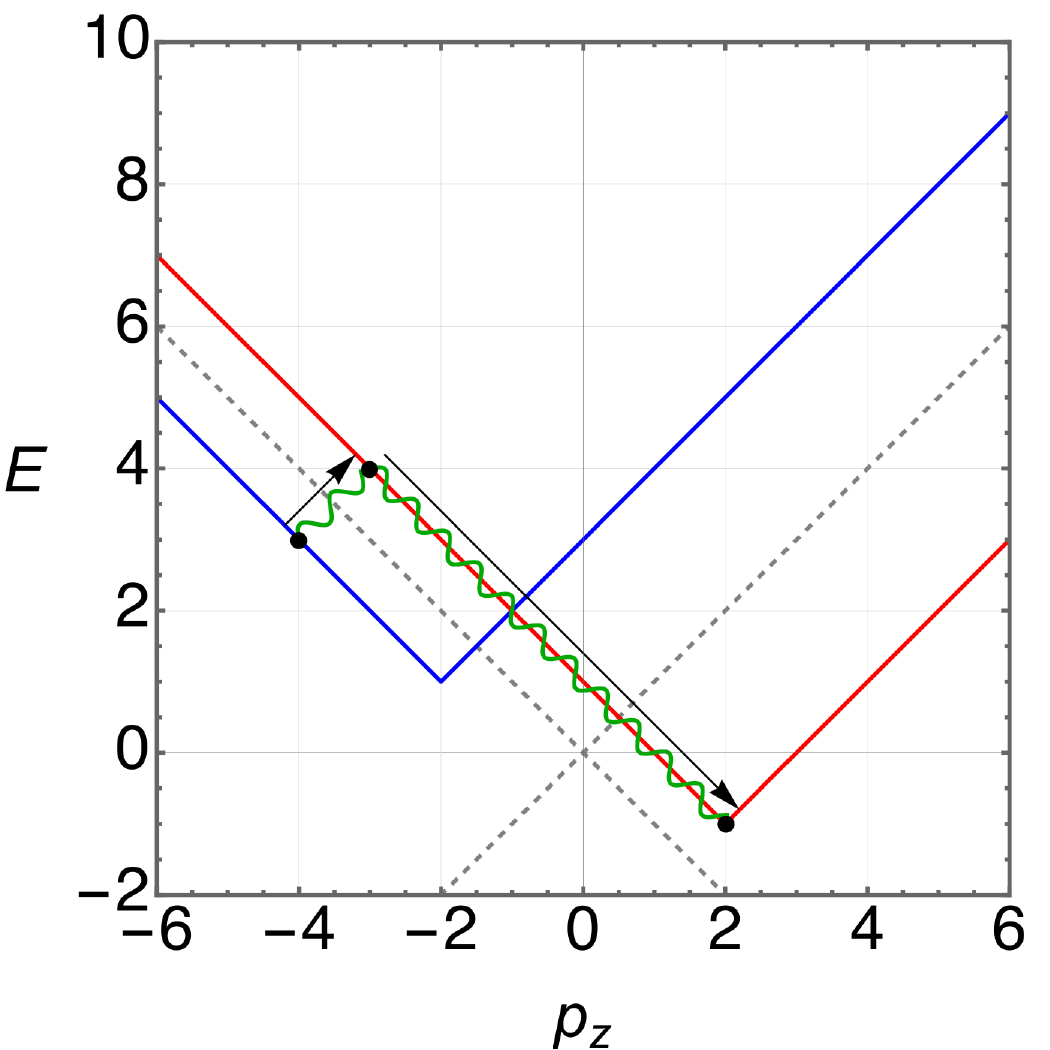}}
\subfigure[]{\label{figb2c}\includegraphics[width=2.2in]{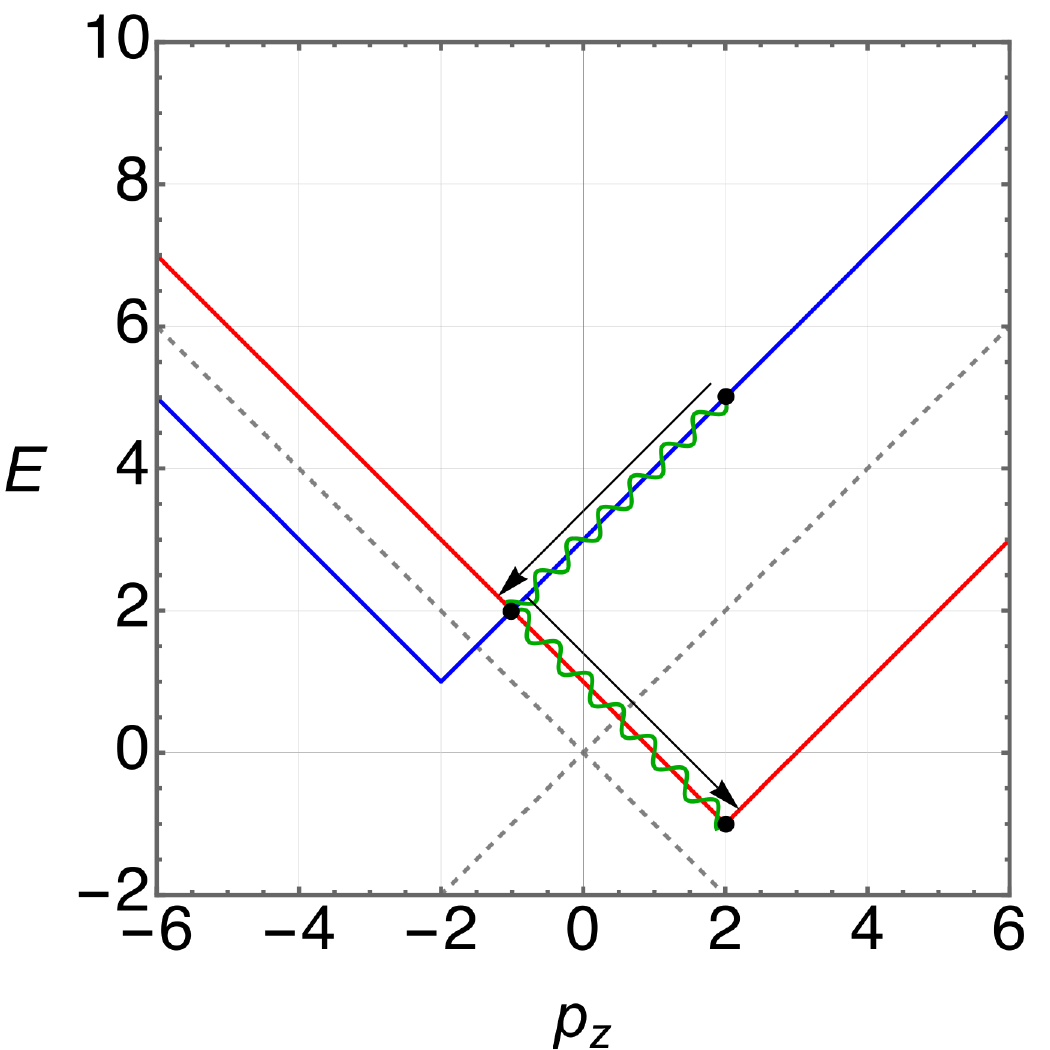}}
\subfigure[]{\label{figb2d}\includegraphics[width=2.2in]{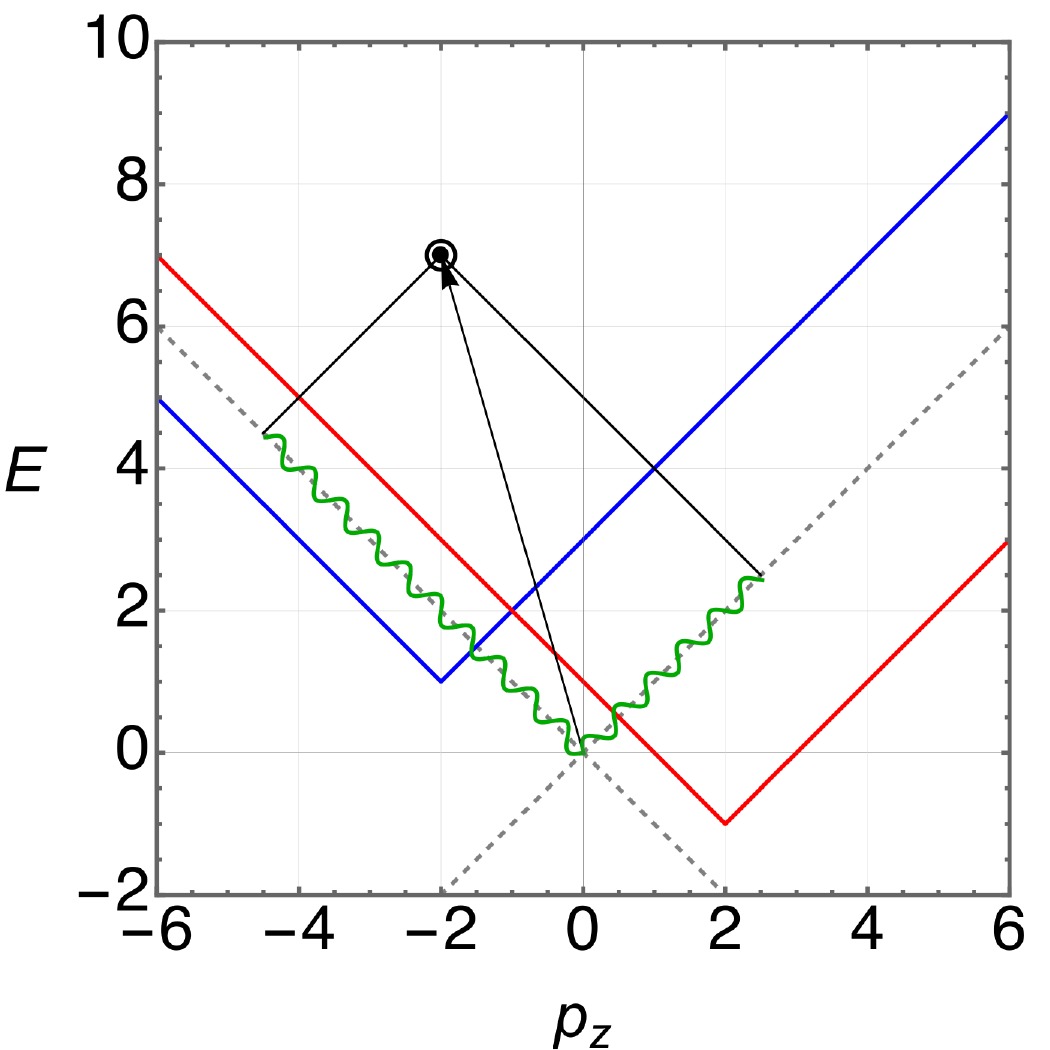}}
\caption{Kinematics of fermion-scalar processes.}
\label{fig:initial-fermion-decay}
\end{figure}

Consider first a fermion with positive energy 
occupying a state on the branch of the dispersion relation
that includes negative energies.
This fermion can settle into a negative-energy state 
via the emission process $f\rightarrow f+\sss$
with parallel or antiparallel fermion 3-momenta,
because the required difference between the initial and final fermion 4-momenta 
is always a 4-vector lying on the scalar lightcone,
and hence the process can always conserve 4-momentum.
An example of this process is shown in Fig.~\ref{figb2a}.
The fermion states are indicated by dots,
while the emission of a scalar is represented by a wavy line. 

In contrast, 
a generic fermion with positive energy occupying a state
on the other branch of the dispersion relation
must undergo a suitable sequence of processes 
involving scalar emission and absorption 
to transfer dispersion branches,
after which it too can settle in a negative-energy state
by scalar emission. 
An example involving absorption of one scalar followed by emission of another
is provided in Fig.~\ref{figb2b},
while an example with two sequential scalar emissions
is displayed in Fig.~\ref{figb2c}.

Finally,
any potential excess positive-energy fermion-antifermion pairs
must annihilate into scalars,
$f\overline{f}\rightarrow n\sss$.
To demonstrate that this requirement 
is consistent with conservation of 4-momentum,
it suffices to consider $n=1$ and $2$.
For $n=2$,
if the initial $f$ and $\overline{f}$ states 
are both situated inside the scalar light cone,
then their momentum sum lies inside the cone
and matches the momentum sum for a two-scalar final state,
and hence the annihilation can take place.
An example of this scenario is shown in Fig.~\ref{figb2d},
where the momentum sum of the initial $f$ and $\overline{f}$ pair
is shown as an arrow.
If instead the initial energy-momentum states 
of either or both of $f$ and $\overline{f}$
fall outside the scalar light cone, 
these states must first be brought inside the cone 
by the absorption $f+\sss \rightarrow f$ of a suitable scalar.
Note that initial configurations exist for which it suffices to lift 
either the fermion or the antifermion to the inside of the scalar light cone. 
For $n=1$,
the single-scalar annihilation $f\overline{f}\rightarrow \sss$ 
can also play a role.
This case is enabled only for specific initial configurations.

\end{document}